\documentclass[aps,prb,reprint, showpacs, superscriptaddress]{revtex4-1}
\usepackage{graphicx}
\usepackage{dcolumn}
\usepackage{bm}
\usepackage{amssymb}
\usepackage{amsmath}
\usepackage{color}
\usepackage{hyperref}
\usepackage[usenames,dvipsnames]{xcolor}
\newcommand{\kp}{\mathbf{k}\cdot\mathbf{p}}

\newcommand{\bk}{\mathbf{k}}
\newcommand{\br}{\mathbf{r}}
\newcommand{\bx}{\mathbf{x}}

\newcommand{\bu}{\mathbf{u}}

\newcommand{\mb}[1]{\mathbf #1}

\newcommand{\ubx}{u_{n0}^*(\mathbf{x}; \varepsilon(\mathbf{x}))}
\newcommand{\ubxp}{u_{n'0}^*(\mathbf{x}; \varepsilon(\mathbf{x}))}
\newcommand{\ubxm}{u_{m0}^*(\mathbf{x}; \varepsilon(\mathbf{x}))}
\newcommand{\ux}{u_{n0}^*(x; \varepsilon(x))}

\hypersetup{
     colorlinks=true,       
     linkcolor=blue,          
     citecolor=blue,        
     filecolor=magenta,      
     urlcolor=black         
}

\begin{document}

\title{Envelope function method for electrons in slowly-varying inhomogeneously deformed crystals}
\author{Wenbin Li}
\affiliation{Department of Materials Science and Engineering,
Massachusetts Institute of Technology, Cambridge, Massachusetts 02139,
USA}
\author{Xiaofeng Qian}
\affiliation{Department of Nuclear Science and Engineering, Massachusetts Institute of Technology, Cambridge, Massachusetts 02139, USA}
\author{Ju Li}
\email{liju@mit.edu}
\affiliation{Department of Nuclear Science and Engineering, Massachusetts Institute of Technology, Cambridge, Massachusetts 02139, USA}
\affiliation{Department of Materials Science and Engineering, Massachusetts Institute of Technology, Cambridge, Massachusetts 02139, USA}


\begin{abstract}
  We develop a new envelope-function formalism to describe electrons
  in slowly-varying inhomogeneously strained semiconductor crystals. A
  coordinate transformation is used to map a deformed crystal back to
  geometrically undeformed structure with deformed crystal
  potential. The single-particle Schr\"{o}dinger equation is solved in
  the undeformed coordinates using envelope function expansion,
  wherein electronic wavefunctions are written in terms of
  \textit{strain-parametrized} Bloch functions modulated by slowly
  varying envelope functions. Adopting local approximation of
  electronic structure, the unknown crystal potential in
  Schr\"{o}dinger equation can be replaced by the strain-parametrized
  Bloch functions and the associated strain-parametrized energy
  eigenvalues, which can be constructed from unit-cell level
  \textit{ab initio} or semi-empirical calculations of homogeneously
  deformed crystals at a chosen crystal momentum. The Schr\"{o}dinger
  equation is then transformed into a coupled differential equation
  for the envelope functions and solved as a generalized matrix
  eigenvector problem. As the envelope functions are slowly varying,
  coarse spatial or Fourier grid can be used to represent the envelope
  functions, enabling the method to treat relatively large systems. We
  demonstrate the effectiveness of this method using a one-dimensional
  model, where we show that the method can achieve high accuracy in
  the calculation of energy eigenstates with relatively low cost
  compared to direct diagonalization of Hamiltonian.  We further
  derive envelope function equations that allow the method to be used
  empirically, in which case certain parameters in the envelope
  function equations will be fitted to experimental data.
\end{abstract}


\maketitle


\section{Introduction}
It has long been recognized that elastic strain can be used to tune
the properties of materials. This idea of elastic strain engineering
(ESE) is straightforward because the derivative of a material property
$P$ with respect to applied elastic strain $\varepsilon$, $\partial
P/ \partial \varepsilon$, is in-general non-zero.\cite{Li14}
However, ESE has traditionally been limited by the small amount of
elastic strain a material can accommodate, before plastic deformation
or fracture occurs. Recent experiments, however, reveal a class of
ultra-strength materials\cite{Li14} whose elastic strain limit can
be significantly higher than conventional bulk solids. Notable
examples are two-dimensional (2D) atomic crystals such as graphene and
monolayer molybdenum disulfide (MoS$_2$).\cite{Novoselov05} The
experimentally measured elastic strain limit of graphene can be as
high as 25\%,\cite{Lee08, Liu07} while that of bulk graphite seldom
reaches 0.1\%. Monolayer MoS$_2$ can also sustain effective in-plane
strain up to 11\%.\cite{Bertolazzi11} Such large elastic strain limit
make it possible to induce significant material property changes
through the application of elastic strain. In particular,
position-dependent properties can be induced by applying
\textit{inhomogeneous} strain which is slowly varying at atomic scale
but has large sample-wide difference. For instance, Feng \textit{et
  al} demonstrated that indenting a suspended MoS$_2$ monolayer can
create a local electronic bandgap profile in the monolayer with $1/r$
spatial variation, $r$ being the distance to the center of indenter
tip.\cite{Feng12} This creates an ``artificial atom'' in which
electrons moves in a semiclassical potential resembling that of a
two-dimensional hydrogenic atom. In this article, we will develop a
new envelope function formalism that could be used to study the
electronic structure of such slowly-varying inhomogeneously strained
crystals.

\textit{Ab initio} electronic structure methods such as density
functional theory (DFT) are nowadays routinely used to calculate the
properties of materials. However, the steep scaling of computational
cost with respect to system size limits their use to periodic solids,
surfaces and small clusters. An inhomogeneously strained structure
usually involves a large number of atoms and thus fall beyond the
current capabilities of these methods.

In the past, several semi-empirical electronic structure methods
capable of treating systems larger than \textit{ab initio} methods
have been developed to study the electrical and optical properties of
semiconductor nanostructures. Among those the most notable are the
empirical tight binding method, \cite{Slater54, DiCarlo03} empirical
pseudopotential method (EPM) \cite{Zunger98, Canning00, Wang99} and
multi-band $\kp$ envelope function method. \cite{Luttinger55, Bastard81, Baraff91, Gershoni93, Burt92, Burt99}
Both tight-binding and EPM
are microscopic methods \cite{DiCarlo03} that treat the electronic
structure at the level of atoms, while the multi-band $\kp$ envelope
function method describes electronic structure at the level of the
envelope of wavefunctions, whose lengthscale is in general much larger
than the lattice constant. Excellent articles discussing the merits
and shortcomings of these methods exist in the
literature. \cite{DiCarlo03, Zunger98} Below we shall briefly review
the multi-band $\kp$ envelope function method and the EPM method, as
these two methods have been demonstrated to treat vary large
nanostructures (up to a million atoms \cite{Wang96, Wang99}) and are
most relevant to our article.

The starting point of wavefunction based semi-empirical electronic
structure methods is usually the single-particle Schr\"{o}dinger
equation:
\begin{equation}
  \left[\frac{\mathbf{p}^2}{2m} + V(\mathbf{r})\right ] \Psi(\mathbf{r}) = E \Psi(\mathbf{r}).
\end{equation}
Here $V(\mathbf{r})$ is the crystal potential; $\Psi(\mathbf{r})$ is
the electronic wavefunction. In $\kp$ envelope function method,
$\Psi(\mathbf{r})$ is expanded in terms of a complete and orthonormal
basis set
$\chi_{n\mb{k}_0}=e^{i\mathbf{k}\cdot\mathbf{r}}\psi_{n\mb{k}_0}(\br)$,
\cite{Luttinger55} where $\psi_{n\mb{k}_0}(\br)$ represent the Bloch
functions of the underlying periodic solid at a reference crystal
momentum $\mb{k}_0$. Mathematically, the expansion is written as
\begin{equation}
  \Psi(\br) = \sum_{n\bk} c_{n\bk} \left\{ e^{i\mb{k}\cdot\mb{r}} \psi_{n\mb{k}_0}(\br) \right\}.
\end{equation}
The summation is over band index $n$ and wave vector $\mathbf{k}$,
which is restricted to the first Brillouin zone (BZ) of the
crystal. This expansion can be re-written as
\begin{equation}
  \Psi(\mathbf{r}) = \sum_{n}\left( \sum_{\bk} c_{n\bk} e^{i\mathbf{k}\cdot\mathbf{r}}\right) \psi_{n\mb{k}_0}(\br)  =\sum_n
  F_n(\br) \psi_{n\mb{k}_0}(\br).
\end{equation}
The functions $F_n(\br) = \sum_{\bk} c_{n\bk}
e^{i\mathbf{k}\cdot\mathbf{r}}$ are called envelope functions because
they are smooth functions at the unit-cell level due to the
restriction of wave vector $\bk$ within the first BZ. If all bands $n$
are kept, the above expansion is complete. In practical calculation,
only a few bands close to Fermi energy are included. The reference
crystal momentum $\mb{k}_0$ is normally chosen to be the wave vector
corresponding to the valence band maximum or conduction band minimum
of a semiconductor.

Using this expansion, the Schr\"{o}dinger equation can be turned into
coupled differential equations for the envelope functions in the
following general form
\begin{equation}
  \sum_n H(\br,\nabla)_{mn}F_n(\br) = EF_m(\br).
\end{equation}
In $\kp$ envelope function method, $H(\br, \nabla)$ is assumed to have
the same form as the $\kp$ Hamiltonian for bulk crystal,\cite{Kane57}
after replacing the momentum operators $k_x$, $k_y$, $k_z$ in $\kp$
Hamiltonian by $-i\partial/\partial x$, $-i\partial/\partial y$, and
$-i\partial/\partial z$.\cite{Luttinger55, Bastard81, Baraff91} The
empirical parameters in $\kp$ Hamiltonian are fitted to the observed
properties of bulk crystals or nanostructures themselves. A Numerical
solution of the coupled differential equations gives energy
eigenvalues and the associated envelope functions. This method has
been successfully applied to semiconductor
superlattice,\cite{Bastard81, Gershoni93} quantum wires,\cite{Stier97}
and quantum dots.\cite{Stier99, Efros00}

The $\kp$ envelope function method can treat the effect of
homogeneous strain by incorporating it as deformation potential.
\cite{Bardeen50, Herring56, Bir74, Gershoni93} Deformation potential
theory assumes small applied strain, such that the strain-induced band
edge shift of bulk crystals can be expanded to first-order in terms of
the applied strain tensor $\varepsilon$, $\Delta E =
\sum_{ij}\Xi_{ij}\varepsilon_{ij}$, where $\Xi_{ij}$ are deformation
potentials. A detailed practical implementation of deformation
potential in $\kp$ envelope function method can be found in
literature.\cite{Gershoni93} Extension of the $\kp$ envelope function
method to inhomogeneous strain was carried out by
Zhang. \cite{Zhang94}

The EPM method \cite{Chelikowsky76, Zunger98, Wang99} is another
well-known approach to nanostructure electronic structure
calculation. EPM solves the single-particle Schr\"{o}dinger equation
non-self-consistently through the use of empirical pseudopotential. In
EPM, the crystal potential $V(\br)$ is represented as a superposition
of screened spherical atomic pseudo-potentials \cite{Zunger98}
\begin{equation}
  V(\br) = \sum_{\mathrm{atom}}v_{\mathrm{atom}} (\br - \mathbf{R}_{\mathrm{atom}}).
\end{equation}
The atomic pseudo-potentials can be extracted from DFT local
density-approximation (LDA) calculations on bulk systems, and then
empirically adjusted to correct the LDA band structure error.
\cite{Wang95} As the laborious self-consistent potential determination
procedures in \textit{ab-initio} calculation are avoided, EPM is
computationally cheaper and faster, enabling it to treat large
nanostructures.\cite{Canning00} Zunger and collaborators showed that
EPM can be more advantageous to $\kp$ method due to its
non-perturbative nature as well as preservation of atomic-level
structural details.\cite{Wood96, Fu98} EPM treats strain effects by
weighting the atomic pseudopotentials with a scalar pre-factor fitted
to observed properties of strained crystals.\cite{Wang99} While EPM is
appealing in many ways, its wide use is limited by the complications
involved in pseudopotential fitting and Hamiltonian diagonalization.

In this article, we develop a new envelope function formalism to solve
the electronic states in slowly-varying inhomogeneously strained
semiconductor crystals. We aim to develop a method that takes
advantage of the numerical efficiency of multi-band $\kp$ envelope
function method, while at the same time incorporates certain
microscopic electronic structure information at the level of
\textit{ab intio} or EPM method. In speaking of a slowly-varying
inhomogeneously strained semiconductor crystal, we mean that the
variation of strain in the crystal is very small over the distance of
a unit cell, but can be quite large sample-wide (more than a few
percent). Our method assumes, with justification, that in such
slowly-varying inhomogeneously strained semiconductors, the local
crystal potential at the unit-cell level can be well approximated by
that of a homogeneously strained crystal with the same strain
magnitude. Hence, significant amount of local electronic structure
information can be obtained from unit-cell level \textit{ab inito} or
EPM calculation of homogeneously strained crystals, which can then be
incorporated into the solution of global electronic structure using
the framework of envelop function method.  To achieve such local to
global electronic structure connection, the global wavefunctions will
be expanded in terms of a small set of Bloch functions
\textit{parametrized} to the strain field $\varepsilon(\bx)$ in the
deformed crystal, each of which is multiplied by a slowly varying
envelope function. The strain-parametrized Bloch functions are
constructed by smoothly connecting the Bloch functions of
homogeneously strained crystals, a process made possible by a
coordinate transformation method that maps the deformed crystal back
to a undeformed one with deformed crystal potential. This set of
strain-parametrized Bloch functions, together with strain-parametrized
energy eigenvalues associated with those Bloch functions, can then be
used to eliminate the unknown crystal potential term in the global
Schr\"{o}inger equation for the inhomogeneously strained crystal. The
electronic structure problem will subsequently be turned into a set of
coupled differential equations for the envelope functions, and solved
as a generalized matrix eigenvector problem. Due to the slowly-varying
nature of the envelope functions, coarse spatial or Fourier grid can
be used to represent them, therefore reduces the computational cost of
the method compared to full-scale \textit{ab initio} or (potentially)
EPM calculation of the inhomogeneously strained crystals.

The structure of the paper is as follows. In Sec.~\ref{sec:formalism},
we lay out the general formalism of our envelope function method for
slowly-varying inhomogeneously strain crystals. To demonstrate its
effectiveness, we will apply the method to a model one-dimensional
strained semiconductor in Sec.~\ref{sec:1DModel}. In
Sec.~\ref{sec:3DIssues}, we will discuss the practical issues when
applying the method to three-dimensional realistic solids. Finally, we
will derive in Sec.~\ref{sec:empirical} a set of differential
eigenvalue equations for the envelope functions when our method is
used as a purely empirical fitting scheme.

\section{Formalism \label{sec:formalism}}


\subsection{Coordinate Transformation}

To facilitate the formulation of our envelope function method for
slowly-varying inhomogeneously strained crystal, we first elaborate a
coordinate transformation method which converts the electronic
structure problem of a deformed crystal to a undeformed one with
deformed crystal potential. This approach has been employed to study
electron-phonon interactions,\cite{Whitfield61} and to prove extended
Cauchy-Born rule for smoothly deformed crystals.\cite{E10, E11,
  E13}. The construction here partly follows E \textit{et al}.
\cite{E13}

In laboratory Cartesian coordinates, denote by $\mathbf{X}_i$ and
$\mathbf{X}'_i$ the position vectors of the $i$-th atom in a crystal
before and after deformation, we can write
\begin{equation}
  \mathbf{X}'_i = \mathbf{X}_i + \mathbf{U}_i.
\end{equation}
Above, $\mathbf{U}_i$ is the displacement of the $i$-th
atom. $\mathbf{U}_i$ is assumed to follow a smooth displacement field
$\bu(\bx)$ in the smoothly deformed crystal, \textit{i.e.}, there
exists a smooth displacement field $\bu(\bx)$, which maps every atom
in the crystal to a new position $\mathbf{X}'_i = \mathbf{X}_i +
\bu(\mathbf{X}_i)$. This assumption is closely related to the
Cauchy-Born rule \cite{Ericksen08} in solid mechanics.

Since the smooth displacement field $\bu(\bx)$ is defined for every
point in the space, it can be used to map a function as well. For
example, if a function $f(\bx)$ is originally defined for an unstrained
crystal, which for example can be the crystal potential $V(\bx)$ or
wavefunction $\Psi(\bx)$, after mapping it becomes a new function
$h(\bx')$ given by
\begin{equation}
 h(\bx + \bu(\bx)) = f(\bx),
\end{equation}
since the value of function $h(\bx')$ at point $\bx' = \bx + \bu(\bx)$
is mapped from function $f(\bx)$ at point $\bx$. This mapping of a known function
defined in a undeformed crystal to a deformed crystal can be done reversely. Suppose,
for example, the crystal potential of a deformed crystal is $V(\bx')$, it can be mapped
back to a function $V^*(\bx)$ defined in the ``undeformed coordinates'' $\bx$ as
\begin{equation}
V^*(\bx) = V(\bx + \bu(\bx)).
\end{equation}
Namely, the value of function $V^*(\bx)$ at position $\bx$ is the same
as the value of function $V(\bx')$ at $\bx'=\bx +
\bu(\bx)$. Hereafter, the appearance of the superscript ``$*$'' on a
function denotes that the function has been mapped back to undeformed
coordinates $\bx$ with the following general mapping rule
\begin{equation}
f^*(\bx) = f(\bx+\bu(\bx)),
\end{equation}
where $f(\bx')$ is a function defined for a deformed crystal.

We can apply the above mapping, which is essentially a nonlinear coordinate
transformation, to differential operators as well, such as the
Hamiltonian operator in the Schr\"{o}dinger equation.  In Hartree
atomic units, the Schr\"{o}dinger equation for deformed crystal reads
\begin{equation}
  \left[-\frac{1}{2}\Delta + V(\bx')\right]\Psi(\bx') = E \Psi(\bx'),
\end{equation}
where $\Delta$ is the Laplacian. Applying the following deformation mapping
(coordinate transformation) to the Schr\"{o}dinger equation,
\begin{equation}
  \bx' = \bx + \bu(\bx),
\end{equation}
it will be transformed into the undeformed coordinates $\bx$ as
\begin{equation}
  \left[-\frac{1}{2}\Delta^* + V^*(\bx)\right] \Psi^*(\bx) = E \Psi^*(\bx).
  \label{eq:Mapped_Schrodinger_Equation}
\end{equation}
$\Delta^*$, $V^*(\bx)$ and $\Psi^*(\bx)$ are the Laplacian, crystal
potential and wavefunctions mapped to undeformed coordinates,
respectively. $\Delta^*$ can be explicitly written out as
\begin{eqnarray}
  \Delta^* &=& \left((I+\nabla \bu)^{-T}\nabla \right)\cdot \left( (I+\nabla \bu)^{-T}\nabla \right) \nonumber \\
  &\equiv& a_{ij}(\bx)\frac{\partial^2}{\partial x_i \partial x_j} + b_i(\bx)\frac{\partial}{\partial x_i},
\label{eq:Mapped_Laplacian}
\end{eqnarray}
where $a_{ij}(\bx)$ and $b_i(\bx)$ are given by
\begin{eqnarray}
  && a_{ij}(\bx) =  \left(I+\nabla \bu(\bx) \right)_{im}^{-1} \left( I+\nabla \bu(\bx) \right)_{mj}^{-T}, \\
  && b_i(\bx)   =   \left( I+\nabla \bu(\bx) \right)_{nm}^{-1} \frac{\partial}{\partial x_n} \left(I+\nabla \bu(\bx) \right)_{mi}^{-T}.
\end{eqnarray}
Above, $\nabla \bu$ is the deformation gradient matrix (field) whose elements
are given by $(\nabla \bu)_{mn} = \partial u_m / \partial x_n$. $I$ is
identity matrix. The superscript $-1$ denotes matrix inversion; $-T$
denotes matrix inversion \textit{and} transposition. Einstein
summation applies when an index is repeated. It can be checked that
when $\bu(\bx) = 0$, \textit{i.e.}, the crystal is undeformed,
$a_{ij}(\bx) = \delta_{ij}$, $b_i(\bx) = 0$, leaving the Laplacian
untransformed.


\subsection{Strain-Parametrized Expansion Basis}
\label{subsec:parametrized_basis}
\begin{figure}[t!]
\includegraphics[width=0.45\textwidth]{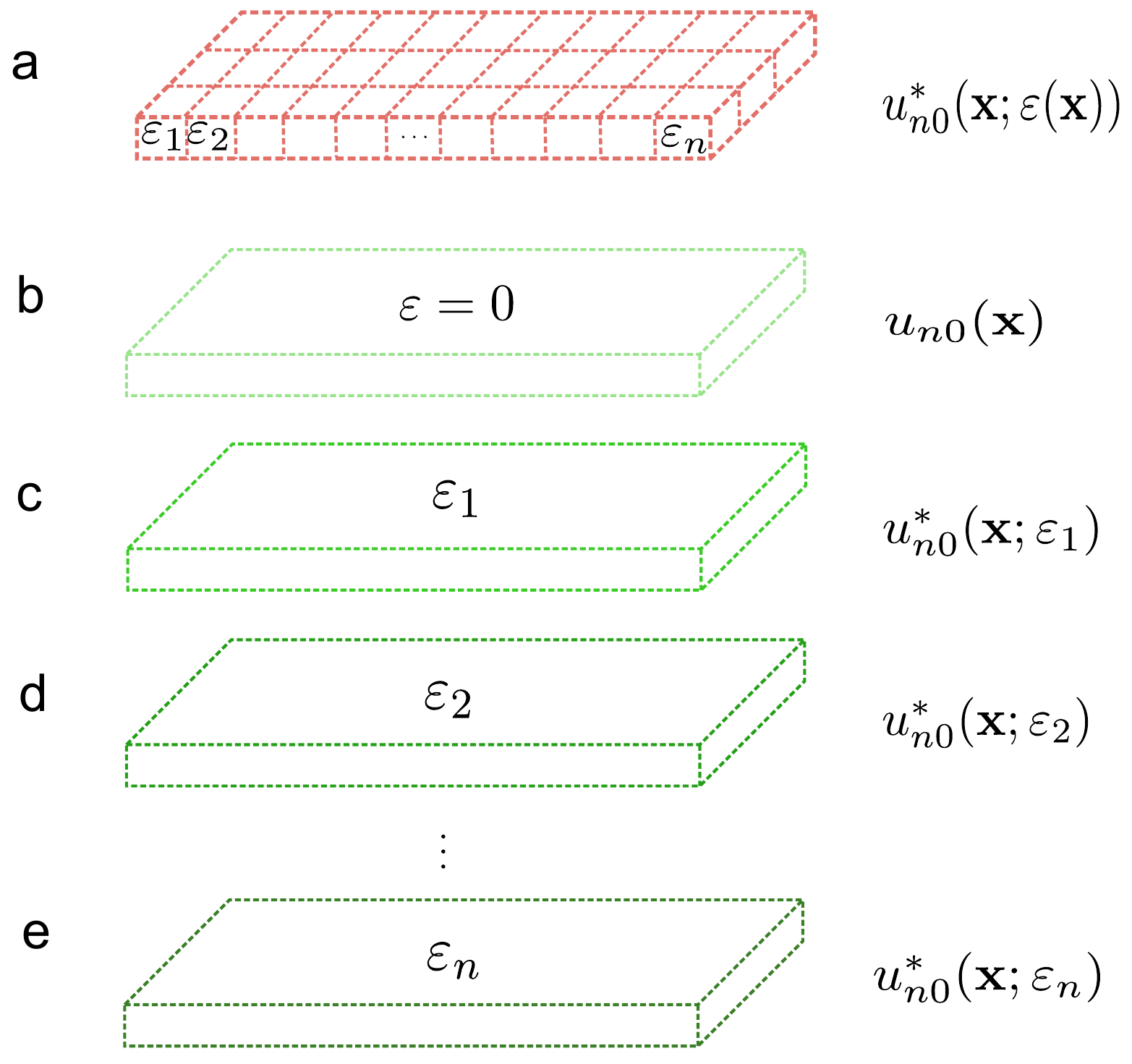}
\caption{\label{fig:fig1} Schematic of strained crystals mapped back
  to undeformed coordinates. After mapping, the atomic coordinates of
  a strained crystal will be the same as those of a undeformed
  crystal, but the crystal potential will be different. (a)
  Inhomogeneously strained crystal. The local strain $\varepsilon_n$
  are labeled. (b) Unstrained crystal. (c-e) Homogeneously strained
  crystals. The mapped Bloch functions $u_{n0}^*(\bx; \varepsilon)$
  are written alongside.}
\end{figure}
To proceed with our envelope function expansion for inhomogeneously
strained crystals, we first imagine a series of $homogeneously$
strained crystals with different strain tensors
$\mathbf{\varepsilon}$, all of which are then mapped back to
undeformed coordinates following the same coordinate transformation
elaborated in the previous section. Fig.~\ref{fig:fig1} illustrates
this procedure. For a homogeneously strained crystal, we can choose
the reference unstrained crystal such that the rotation component of
the displacement field is zero, which allow the displacement
$\bu(\bx)$ to be be written as $\bu(\bx) = \varepsilon \bx $, namely
$u_{i} = \varepsilon_{ik}x_k$. It then follows from
Eq.~\ref{eq:Mapped_Schrodinger_Equation} and
Eq.~\ref{eq:Mapped_Laplacian} that the Schr\"{o}dinger equation for
homogeneously strained crystals transforms into undeformed coordinates
as
\begin{equation}
\left[-\frac{1}{2}(I+\varepsilon)^{-1}_{im}(I+\varepsilon)^{-T}_{mj}\frac{\partial^2}{\partial x_i\partial x_j}
+ U^*(\bx; \varepsilon) \right] \Psi^* = E\Psi^*,
\end{equation}
Here, to distinguish the crystal potential of homogeneously strained
crystal from that of inhomogeneously strained crystal, we have used
the symbol $U^*(\bx; \varepsilon)$ to represent the mapped crystal
potential of homogeneously strained crystal with strain tensor
$\varepsilon$. From now on, $U$ and $V$ will be used to represent the
crystal potentials of homogeneously strained crystals and
inhomogeneously strained crystals respectively.

Given a reference crystal momentum $\bk_0$ for unstrained crystal, for
each of the homogeneously strained crystal with strain tensor
$\varepsilon$, their Bloch functions at the corresponding strained crystal
momentum $\mb{k} = (I+\varepsilon)^{-T}\mb{k}_0$ can be written as
\begin{equation}
\psi_{n\bk}(\bx';\varepsilon) = e^{i\bk\cdot\bx'}u_{n\bk}(\bx';\varepsilon).
\end{equation}
$\psi_{n\bk}(\bx';\varepsilon)$ can then be mapped back to undeformed
coordinates and denoted by
\begin{equation}
\psi_{n\bk_0}^*(\bx;\varepsilon) = e^{i\bk_0\cdot\bx}u_{n\bk_0}^*(\bx;\varepsilon).
\end{equation}
Without loss of generality, hereafter we chose the reference crystal
momentum $\bk_0 = 0$, in which case only the periodic part of the
Bloch functions $u_{n0}^*(\bx; \varepsilon)$ will be retained.

For any value of strain $\varepsilon$, the mapped Bloch functions
$u_{n0}^*(\bx; \varepsilon)$ are periodic functions of the original,
undeformed lattice translation vectors. Therefore, each of them can be
expanded in undeformed coordinates in terms of a complete and
orthonormal basis set $\varphi_m(\bx)$, which for example can be plane
waves:
\begin{equation}
  u_{n0}^*(\bx; \varepsilon) = \sum_m C_m^n(\varepsilon)\varphi_m(\bx).
\label{eq:Strained_Blochwave_Expansion}
\end{equation}
The expansion coefficients $C_m^n(\varepsilon)$ will be dependent of
the strain value $\varepsilon$. After this expansion, the strain
$\varepsilon$ dependence of the Bloch functions $u_{n0}^*(\bx;
\varepsilon)$ is separated into the expansion coefficients
$C_m^n(\varepsilon)$. In the absence of strain-induced phase
transition, and choosing the same gauge \cite{Marzari12} for different
strained Bloch functions, these expansion coefficients should be
continuous functions of strain $\varepsilon$. We then can define a
\textit{strain-parametrized} basis set $u_{n0}^*(\bx;
\varepsilon(\bx)) \equiv u_{n0}^*(\bx; \varepsilon =
\varepsilon(\bx))$, which means that at position $\bx$, the values of
the expansion coefficients $C_m^n$ in
Eq.~\ref{eq:Strained_Blochwave_Expansion} take the values of
$C_m^n(\varepsilon = \varepsilon(\bx))$. Mathematically, this can be
written out as
\begin{equation}
\ubx = \sum_m C_m^n\left(\varepsilon(\bx)\right)\varphi_m(\bx),
\label{eq:Parametrized_Expansion_Basis}
\end{equation}
$\ubx$ are named ``strain-parametrized Bloch functions'', since they
are parametrized to the strain field $\varepsilon(\bx)$ in an
inhomogeneous strained crystal. In analogy with the conventional
envelope function method,\cite{Luttinger55} we can use $\ubx$ to
expand the mapped global wavefunctions $\Psi^*(\bx)$ of
inhomogeneously strained crystals,
\begin{equation}
  \Psi^*(\bx) = \sum_n F_n(\bx) \ubx.
\label{eq:Envelope_Function_Expansion}
\end{equation}
The summation is over the band index $n$, but will normally be
truncated to include only bands within certain energy range of
interest, as far-away bands have smaller contributions to the
electronic states under consideration. Here, $F_n(\bx)$ are, as in
conventional envelope function method, considered to be smooth
functions on unit-cell scale. This envelope expansion is, in essence,
a continuous generalization of Bastard's envelope expansion method for
semiconductor heterostructures,\cite{Bastard81} where the
wavefunctions in the barrier and well regions of heterostructures are
expanded in the Bloch functions of respective region.


\subsection{Local Approximation of Strained Crystal Potential and Envelope Function Equation \label{subsec:locality}}
Our next natural step is to substitute $\Psi^*(\bx)$ into the mapped
Sch\"{o}dinger equation for inhomogeneously strained crystal, the
Eq.~\ref{eq:Mapped_Schrodinger_Equation}, which for convenience is
rewritten here as
\begin{equation}
\left[\mathcal{P}^* + V^*(\bx)\right] \Psi^*(\bx) = E \Psi^*(\bx),
\end{equation}
where $\mathcal{P}^*$ is an operator given by
\begin{equation}
  \mathcal{P}^*= -\frac{1}{2}\left[a_{ij}(\bx)\frac{\partial ^2}{\partial x_i \partial x_j}
+ b_i(\bx)\frac{\partial}{\partial x_i} \right].
  \label{eq:operator_P}
\end{equation}
$a_{ij}(\bx)$ and $b_i(\bx)$ have been defined earlier. Replacing
$\Psi^*(\bx)$ by the strain-parametrized envelope function expansion
in Eq.~\ref{eq:Envelope_Function_Expansion}, the above Schr\"{o}dinger equation becomes
\begin{equation}
\begin{split}
  \sum_n \mathcal{P}^*\left[F_n(\bx) \ubx\right] & + \sum_n F_n(\bx) \left[ V^*(\bx)
    \ubx \right] \\&= E\sum_n F_n(\bx) \ubx.
\end{split}
\label{eq:Expanded_Schrodinger}
\end{equation}
The potential energy operator $V^*(\bx)$ is the unknown term in the
Hamiltonian, which in \textit{ab initio} calculation is determined
self-consistently. As we have argued earlier, such self-consistent
calculation of $V^*(\bx)$ is usually impractical for an
inhomogeneously strained crystal due to the large system size. Hence,
we introduce here an important approximation in our method: for a
slowly-varying inhomogeneously strained semiconductor, the crystal
potential $V^*(\bx)$ at position $\bx$ can be well approximated by
that of a homogeneously deformed crystal with same strain tensor
$\varepsilon(\bx)$, if (a) the applied elastic strain field
$\varepsilon(\bx)$ is sufficiently slowly-varying at atomic scale and (b)
long-range electrostatic effects \cite{VandeWalle89} are negligible.
This locality principle for the electronic structure of
insulators/semiconductors has been proved by E \textit{et
  al}. \cite{E10, E11, E13} It is also implicitly implied in the
treatment of strain in the EPM method.  \cite{Wang99} We also note
that, the limitation of local approximation to the effective
potential in the envelope function method has been studied
by other authors \cite{Milanovic82, Foreman07, Yoo10,
  FloresGodoy13}.

Mathematically, the locality principle translates into $V^*(\bx) =
U^*(\bx; \varepsilon(\bx)) + \mathcal{O}(b|\nabla \varepsilon(\bx)|)$,
where $U^*(\bx; \varepsilon(\bx))$ is the strain-parametrized crystal
potential of homogeneously strained crystals, $b$ is the average
magnitude of lattice constants, and $\nabla \varepsilon(\bx)$ is the
gradient of strain field. $b|\nabla \varepsilon(\bx)|$ is thus a
measure of how fast strain varies at atomic scale. Clearly, the
smaller this measure, the better the locality approximation will
be. In the case $\varepsilon(\bx)$ goes to zero, the approximation
becomes exact. Since we are concerned with slowly-varying
inhomogeneously strained crystals in this article, in what follows we
will only keep the term $U^*(\bx; \varepsilon(\bx))$, which is the
zeroth-order term in strain gradient, or the first-order term in
displacement gradient.

Adopting this locality principle greatly facilitates the solution of
the electronic structure problem, as we can now use the local
electronic structure information of homogeneously deformed crystals,
obtained from unit-cell level \textit{ab initio} or semi-empirical
calculations, to eliminate the unknown crystal potential term
$V^*(\bx)$ in Eq.~\ref{eq:Expanded_Schrodinger}. Specifically, we can
write down the local Schr\"{o}dinger equation for the
strain-parametrized expansion basis
\begin{equation}
  \left[\mathcal{P}_0^* + U^*(\bx;\varepsilon(\bx))\right] \ubx
  = \epsilon_{n0}(\varepsilon(\bx)) \, \ubx,
\label{Local_Schrodinger}
\end{equation}
with $\mathcal{P}_0^*$ being
\begin{equation}
  \mathcal{P}_0^* = -\frac{1}{2}(I+\varepsilon(\bx))^{-1}_{im}(I+\varepsilon(\bx))^{-T}_{mj}
  \left.\frac{\partial ^2}{\partial x_i \partial x_j}\right|_{\varphi(\bx)}.
  \label{eq:operator_P0}
\end{equation}
In Eq.~\ref{Local_Schrodinger}, $\epsilon_{n0}(\varepsilon(\bx))$ is
the strain-parametrized energy eigenvalues for band $n$ at the
reference crystal momentum, defined as
$\epsilon_{n0}(\varepsilon(\bx)) \equiv \epsilon_{n0}(\varepsilon =
\varepsilon(\bx))$. The subscript $\varphi(\bx)$ in
$\mathcal{P}_0^*$ denotes that, when the partial derivatives operate
on the strain-parametrized expansion $\ubx = \sum_m
C_m^n\left(\varepsilon(\bx)\right)\varphi_m(\bx) $, they act on
the position dependence coming from $\varphi(\bx)$, but not on the $\bx$
dependence coming from $C_m^n(\varepsilon(\bx))$. To better understand
Eq.~\ref{Local_Schrodinger}, one can look at the limit when the strain
field $\varepsilon(\bx)$ is uniform throughout the
crystal. Eq.~\ref{Local_Schrodinger} then simply becomes a normal
Schr\"{o}dinger equation for homogeneously strained crystal mapped to
undeformed coordinates.

We will now use the local Schr\"{o}dinger equation to eliminate the
potential energy operator $V^*(\bx)$ in global Schr\"{o}dinger
equation. Rearranging Eq.~\ref{Local_Schrodinger}, we have
\begin{equation}
  U^*(\bx;\varepsilon(\bx)) \ubx = \left[-\mathcal{P}_0^* + \epsilon_{n0}(\varepsilon(\bx)) \right]\ubx.
\label{eq:Rearranged_Local_Schrodinger}
\end{equation}
We then replace $V^*(\bx)$ in Eq.~\ref{eq:Expanded_Schrodinger} by
$U^*(\bx;\varepsilon(\bx))$ based on the locality principle, and
replace $U^*(\bx;\varepsilon(\bx)) \ubx$ by the right-hand side of
Eq.~ \ref{eq:Rearranged_Local_Schrodinger}.  Finally, we reach the
following coupled differential equation for the envelope functions
$F_n(\bx)$:
\begin{equation}
\begin{split}
  \sum_n\mathcal{P}^*&\left[F_n(\bx) \ubx \right] - \sum_n F_n(\bx) \mathcal{P}_0^*\left[\ubx\right] \\
  &=\sum_n F_n(\bx) \left[ E - \epsilon_{n0}(\varepsilon(\bx)) \right]\ubx.
\end{split}
\label{eq:Coupled_Envelope_Function_Equation}
\end{equation}
This coupled differential eigenvalue equation is the central equation
we need to solve in our envelope function method. The unknowns in the
equations are the global energy eigenvalues $E$ and their associated
envelope functions $F_n(\bx)$. The strain-parametrized Bloch functions
$\ubx$ and their energy eigenvalues $\epsilon_{n0}(\varepsilon(\bx))$,
can be constructed using \textit{ab inito} or semi-empirical
calculation of homogeneously strained crystals at unit-cell level,
using the procedures described in
Sec.~\ref{subsec:parametrized_basis}. The coupled differential
equation can be solved numerically by expanding the envelope functions
in an appropriate basis, and then turned into a matrix eigenvalue
equation. The expansion basis can be judiciously chosen to reflect the
symmetries that the envelope functions could have. The most general
expansion basis, however, are plane waves:
\begin{equation}
  F_n(\bx) = \sum_{\bk} B_{n\bk} e^{-i\bk \cdot \bx}.
\label{eq:EnvelopeFunction_PlaneWave_Expansion}
\end{equation}
Plugging the above equation into
Eq.~\ref{eq:Coupled_Envelope_Function_Equation}, it will turn into the
following equation
\begin{widetext}
\begin{equation}
  \sum_n\sum_{\bk} B_{n\bk} \left\{ \mathcal{P}^* \left[ e^{i\bk \cdot \bx} \ubx \right] - e^{i\bk \cdot \bx} \mathcal{P}_0^* \left[ \ubx \right] \right\} = \sum_n \sum_{\bk} B_{n\bk} e^{i\bk \cdot \bx} \left( E - \epsilon_{n0}(\varepsilon(\bx)) \right) \ubx.
\label{eq:Coupled_Envelope_Function_Equation_Expanded}
\end{equation}
\end{widetext}
We then multiply the both sides of
Eq.~\ref{eq:Coupled_Envelope_Function_Equation_Expanded} by
$\left[e^{i\bk' \cdot \bx}u_{m0}^*(\bx; \varepsilon(\bx))
\right]^\dagger$ (dagger denotes complex conjugation), and then
integrate both side over the whole crystal volume $V$. This results in
$N\times M_{\bk}$ independent linear equations, where $N$ is the
number of bands included in the envelope function expansion in
Eq.~\ref{eq:Envelope_Function_Expansion}, $M_{\bk}$ is the number of
plane waves used to expand the envelope functions $F_n(\bx)$ in
Eq.~\ref{eq:EnvelopeFunction_PlaneWave_Expansion}. The system of
linear equations are written below as:
\begin{equation}
  \sum_{n\bk} B_{n\bk} \, (W_{n\bk}^{m\bk'} - R_{n\bk}^{m\bk'} + S_{n\bk}^{m\bk'}) = E \sum_{n\bk} B_{n\bk} \, T_{n\bk}^{m\bk'},
\label{eq:MatrixEquation}
\end{equation}
where
\begin{eqnarray}
  &&W_{n\bk}^{m\bk'} = \int_{V} d\bx \ \left[ e^{i\bk' \cdot \bx}u_{m0}^* \right]^\dagger \mathcal{P}^* \left[ e^{i\bk \cdot \bx} u_{n0}^* \right], \nonumber \\
  &&R_{n\bk}^{m\bk'} = \int_{V} d\bx \ e^{i(\bk-\bk') \cdot \bx} (u_{m0}^*)^\dagger \mathcal{P}_0^* u_{n0}^*, \nonumber \\
  &&S_{n\bk}^{m\bk'} = \int_{V} d\bx \ e^{i(\bk-\bk') \cdot \bx} (u_{m0}^*)^\dagger u_{n0}^* \nonumber, \\
  &&T_{n\bk}^{m\bk'} = \int_{V} d\bx \ e^{i(\bk-\bk') \cdot \bx} \epsilon_{n0} \left(\varepsilon(\bx)\right) (u_{m0}^*)^\dagger u_{n0}^* . \nonumber
\end{eqnarray}
$u_{n0}^*$ is short for $\ubx$. The system of linear equations can be
solved numerically as a generalized eigenvector problem to obtain the
eigenvalues $E$ and eigenvectors $B_{n\bk}$.

\section{Application to One-Dimensional Models \label{sec:1DModel}}


\subsection{General Framework \label{sec:1DFramework}}
To demonstrate the effectiveness of our envelope function method, we
will apply the method to one-dimensional (1D) inhomogeneously strained
crystals. We will first lay out the general mathematical framework of
the method in 1D, followed by a specific example in the next
section. Most equations in this section are just 1D special cases of
equations in the previous section.

Suppose a slowly varying inhomogeneous strain $\varepsilon(x)$ is
imposed on a 1D crystal. The strain field corresponds to a
displacement field $u(x) = \int^x \varepsilon(x')dx'$. The operators
$\mathcal{P}^*$ and $\mathcal{P}_0^*$ defined in the previous section
will have the following form
\begin{eqnarray}
&& \mathcal{P}^*=  - \frac{1}{2[1+\varepsilon(x)]^2}\frac{d^2}{d x^2}
                  +\frac{\varepsilon'(x)}{2 \left[1+\varepsilon(x)\right]^3}\frac{d}{d x}, \label{eq:OperatorP_1D}  \\
&& \mathcal{P}_0^* =  -\frac{1}{2[1+\varepsilon(x)]^2}\frac{\partial^2}{\partial x^2}, \label{eq:OperatorP0_1D}
\end{eqnarray}
where $\varepsilon'(x)$ denotes the derivative of strain with respect
to $x$. The partial derivative in $\mathcal{P}_0^*$ implies that, for
a strain parametrized function $f(x;\varepsilon(x))$, the derivative
will not act on the $x$ dependence coming from $\varepsilon(x)$.

The Schr\"{o}dinger equation mapped back to undeformed coordinates
will be
\begin{equation}
\left[\mathcal{P}^* + V^*(x)\right] \Psi^*(x) = E\Psi^*(x),
\label{eq:Mapped_Schrodinger_1D}
\end{equation}
where $V^*(x)$ and $\Psi^*(x)$ are mapped crystal potential and energy
eigenfunction in undeformed coordinates. $E$ is energy
eigenvalue. $\Psi^*(x)$ will be expanded in terms of envelope functions
and strain-parametrized Bloch functions:
\begin{equation}
  \Psi^*(x) = \sum_n F_n(x) \ux
\end{equation}
The strain parametrized Bloch functions $\ux$ satisfies the local Schr\"{o}dinger equation for homogeneously strained crystal
\begin{equation}
  \left[\mathcal{P}_0^* + U^*(x;\varepsilon(x))\right] \ux = \epsilon_{n0}(\varepsilon(x)) \ux.
\label{eq:Local_Schrodinger_1D}
\end{equation}
We then adopt the local approximation of crystal potential
$V^*(x)\approx U^*(x;\varepsilon(x))$, which allows us to use the
above local Schr\"{o}dinger equation to transform
Eq.~\ref{eq:Mapped_Schrodinger_1D} into the following envelope
function equation
\begin{equation}
\begin{split}
  \sum_n &\left\{\mathcal{P}^*\left[F_n(x) \ux \right] - F_n(x) \mathcal{P}_0^*\left[\ux\right]\right\} \\
  &=\sum_n F_n(x) \left[ E - \epsilon_{n0}(\varepsilon(x)) \right]
  \ux.
\end{split}
\end{equation}
Using the explicit forms of $\mathcal{P}^*$ and $\mathcal{P}_0^*$ in
Eq.~\ref{eq:OperatorP_1D} and Eq.~\ref{eq:OperatorP0_1D}, the above
equation can be further written as
\begin{equation}
  \sum_n\left[p_n(x) F_n'' + q_n(x) F_n' + g_n(x) F_n \right] = E \sum_n h_n(x) F_n,
\label{eq:Envelope_Function_1D}
\end{equation}
with $p_n(x)$, $q_n(x)$, $g_n(x)$ and $h_n(x)$ given by
\begin{equation}
\begin{split}
 p_n(x) = &\ux \\
 q_n(x) = &2\frac{d}{dx} \ux - \frac{\varepsilon'(x)}{1+\varepsilon(x)}\ux \\
 g_n(x) = & \frac{d^2}{d x^2}\ux - \frac{\partial^2}{\partial x^2} \ux \\
          &   -  \frac{\varepsilon'(x)}{1+\varepsilon(x)}\frac{d}{dx} \ux \\
          &   -2[1+\varepsilon(x)]^2 \epsilon_{n0}(\varepsilon(x)) \ux \\
 h_n(x) = & -2[1+\varepsilon(x)]^2\ux.
\end{split}
\end{equation}
After constructing the strain parametrized Bloch functions $\ux$ and
$\epsilon_{n0}(\varepsilon(x))$ through unit-cell level calculations of
homogeneously strained crystals and strain-parametrization, described
in Sec.~\ref{subsec:parametrized_basis}, the coupled differential
eigenvalue equation Eq.~\ref{eq:Envelope_Function_1D} can be solved
numerically using the method described in the previous section.


\subsection{Example}
\begin{figure}[t!]
\includegraphics[width=0.35\textwidth]{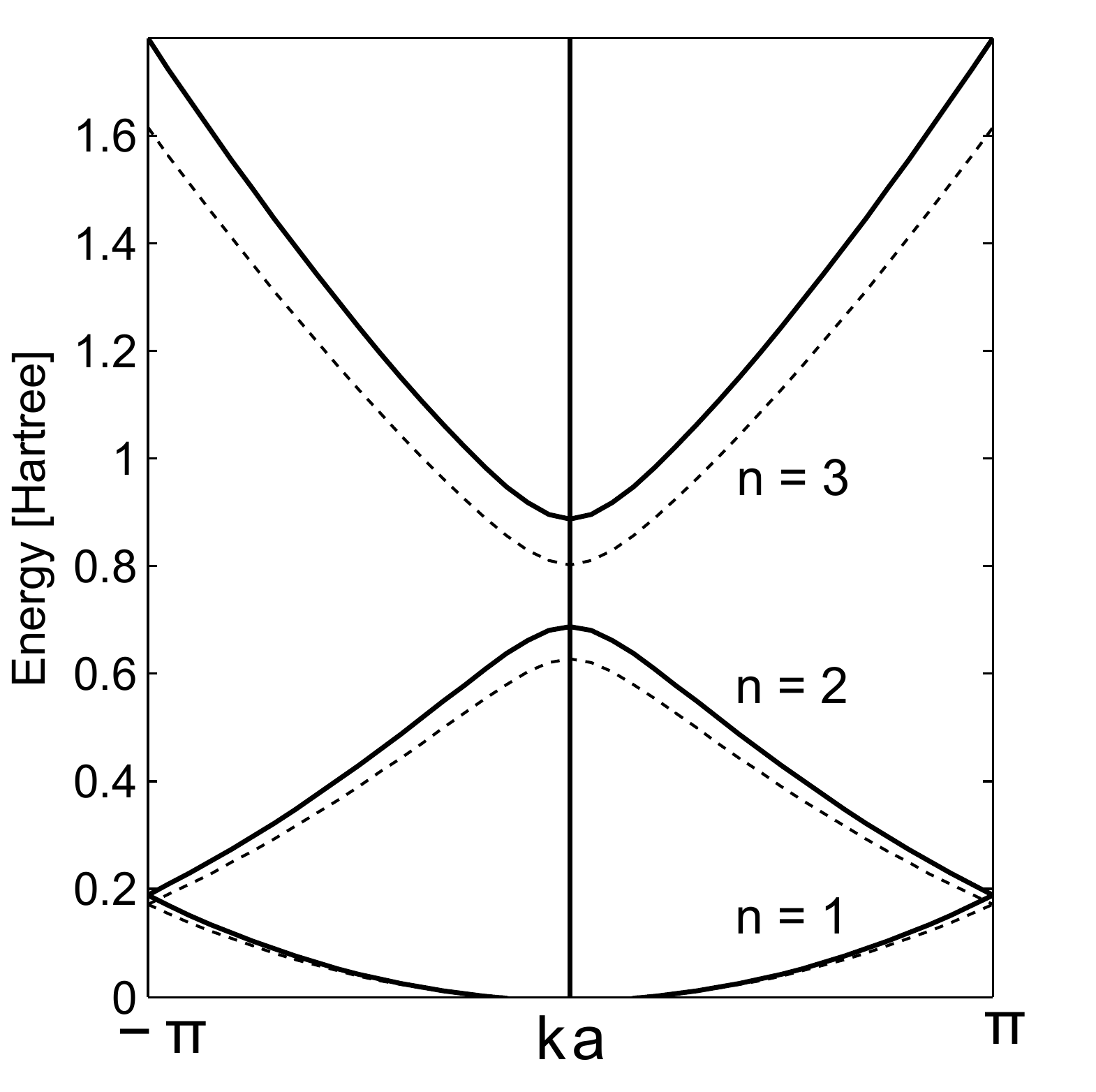}
\caption{\label{fig:fig2} Calculated energy band structure of the
  model 1D crystal before and after applying homogeneous strain (see
  main text for details of the 1D crystal). Only the first three bands
  are presented. Solid and dashed line denote the first three energy
  bands of unstrained crystal and homogeneously strained crystal with
  $\varepsilon = 0.05$, respectively. The axis label $ka$ denotes the
  product of crystal momentum $k$ and lattice constant $a =
  a_0(1+\varepsilon)$.}
\end{figure}

Consider a 1D crystal with lattice constant $a_0$ and the following model crystal
potential
\begin{equation}
  U(x) = -U_0 \cos\left(\frac{4\pi}{a_0}x \right).
\end{equation}
This crystal potential has the following attractive features:

(1) A direct bandgap of magnitude $E_g \approx U_0$ will open up
between the second and third energy band at crystal momentum $k = 0$,
as shown in Fig.~\ref{fig:fig2}. Assuming that the first and second
band are completely filled by electrons while those bands above are
empty, the 1D crystal corresponds to a direct bandgap semiconductor
for which the second band ($n=2$) is the ``valence band'' and the
third band ($n=3$) is the ``conduction band''. We will use this
designation from now on. The bandgap is $E_g\equiv E_c - E_v \approx
U_0$, where $E_c$ is the energy of the conduction band minimum, and $E_v$
is the energy of the valence band maximum.

(2) If we fix $U_0$, the value of bandgap $E_g$ will almost have no
change even when the lattice constant $a_0$ is varied. The change of
lattice constant $a_0$ is natural when we apply strain $\varepsilon$
to the system, namely $a_0$ becomes $a_0(1+\varepsilon)$. While $E_g\equiv E_c
- E_v \approx U_0$ does not change when lattice constant $a_0$ is
changed, the absolute energy values of the conduction band edge $E_c$
and valence band edge $E_v$, however, do shift, mainly due to the
change of kinetic energies for electrons in the system when enlarging
or shrinking the crystal. we can therefore model the strain-induced
energy level shifting without incurring bandgap change in this model
crystal potential.

(3) If we want to model bandgap change when strain is applied, we can
simply write $U_0$ as a function of strain $\varepsilon$. For example,
to model the linear change of bandgap as a function of strain, we can
write $U_0(\varepsilon) = U_0 + K\varepsilon$, where $K$ denotes the
rate of bandgap change as a function of strain.

Hence, the 1D crystal potential is an excellent model system for 1D
semiconductor, whose band edge energy levels $(E_c, E_v)$ and bandgap
$E_g$ can be independently tuned. The crystal potential can therefore
model deformation potential \cite{Bardeen50} while being
mathematically simple and transparent.

In the spirit of the above discussion, we now assume that, after
applying homogeneous strain $\varepsilon$ to the model 1D
semiconductor, its crystal potential has the following form:
\begin{equation}
  U(x';\varepsilon) = -(U_0+K\varepsilon)\cos\left[\frac{4\pi}{a_0(1+\varepsilon)}x'\right].
\label{eq:Potential_HomoStrain_1D}
\end{equation}
This implies that both the energy levels and bandgap of the 1D crystal
will change after applying strain. Comparison of the band-structures
of the 1D crystal before and after deformation for a specific set of
parameters $U_0 = 0.2$, $K = -0.5$, and $\varepsilon = 0.05$ is shown
in Fig.~\ref{fig:fig2}.

The crystal potential of homogeneously strained 1D crystal,
$U(x';\varepsilon)$, is up to now defined in strained coordinates
$x'$. As discussed earlier, we can map the strained crystal potential
back to undeformed coordinates $x$ based the mapping relationship
$x' = x + u(x) = (1+\varepsilon)x$. The mapped crystal potential
becomes the following
\begin{equation}
U^*(x;\varepsilon) = -(U_0+K\varepsilon)\cos(4\pi x/a_0).
\label{eq:Potential_HomoStrain_1D_mapped}
\end{equation}

Suppose now a continuous strain distribution $\varepsilon(x)$ is
defined in the $x$ coordinates. We can define a strain-parametrized
crystal potential $U^*(x;\varepsilon(x))$ such that at position $x$,
we first calculate the strain $\varepsilon(x)$ at $x$, then assign
$U^*(x;\varepsilon(x))$ a value equal to $U^*(x;\varepsilon =
\varepsilon(x))$. Namely,
\begin{equation}
\begin{split}
U^*(x;\varepsilon(x))&\equiv U^*(x; \varepsilon=\varepsilon(x))\\
                     & =  -(U_0+K\varepsilon(x))\cos(4\pi x/a_0).
\end{split}
\end{equation}

\begin{figure}[t!]
\includegraphics[width=0.5\textwidth]{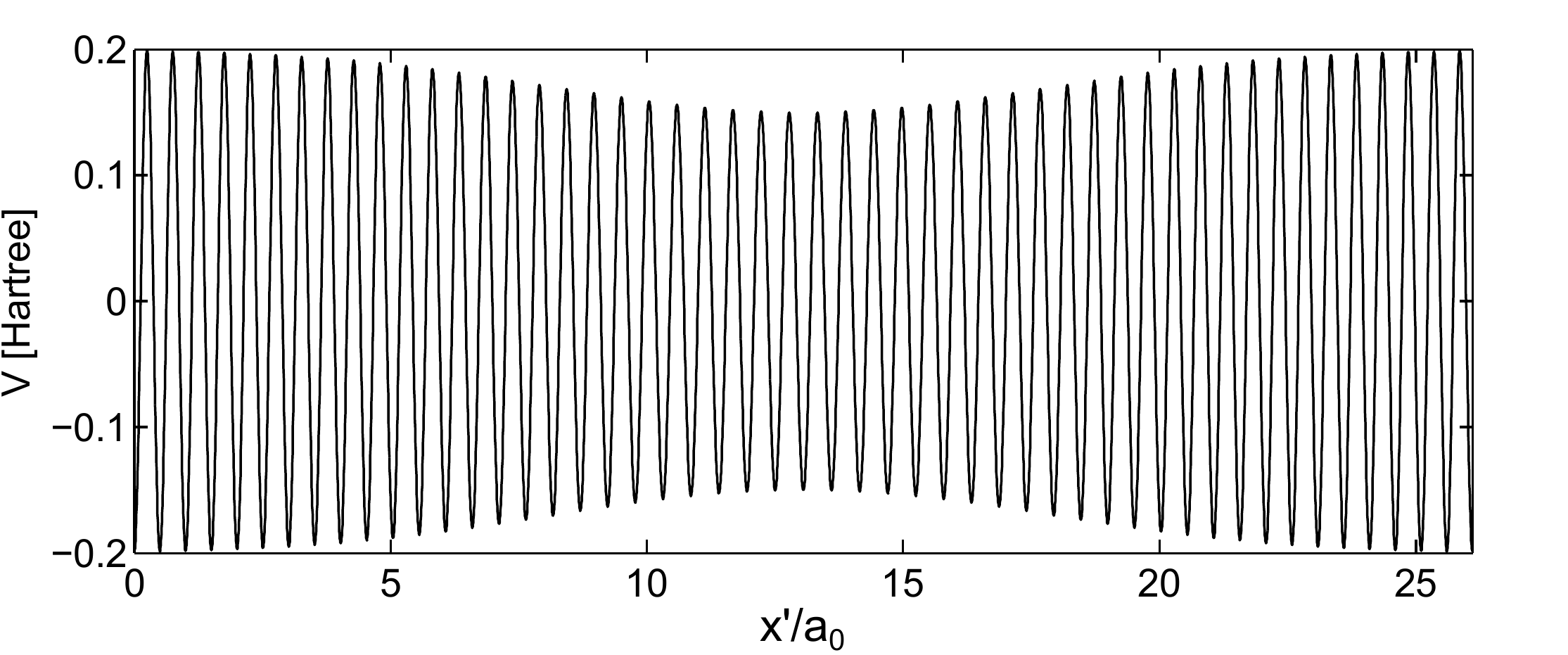}
\caption{\label{fig:fig3} Potential of the model 1D crystal after
  applying Gaussian-type inhomogeneous strain. $V(x') =
  U^*(x;\varepsilon(x)) = -\left[U_0+K\varepsilon(x)\right] \cos(4\pi
  x/a_0)$. $x$ is related to $x'$ via $x' = x + \frac{\sqrt{\pi}
    L}{8}\varepsilon_{\rm max} \left[\mathrm{erf}
    \left(\frac{x-L/2}{L/4}\right) - \mathrm{erf}(-2) \right]$. The
  values of model parameters $U_0$, $K$, $L$ and $\varepsilon_{\rm
    max}$ are given by $U_0 = 0.2$, $K = -0.5$, $ L = 25a_0$ and
  $\varepsilon_{\rm max} = 0.1$.}
\end{figure}

With the above model set-up, we now apply a Gaussian-type
inhomogeneous strain on the 1D crystal. The strain
distribution is given by
\begin{equation}
\varepsilon(x) = \varepsilon_{\rm max} \exp\left[-\frac{(x-L/2)^2}{(L/4)^2}\right],
\label{eq:Inhomo_strain_field_1D}
\end{equation}
where $\varepsilon_{\rm max}$ is the maximum strain value in the
strain field $\varepsilon(x)$, occurring at $x = L/2$. $L$ denotes the
size of crystal.  After applying the inhomogeneous strain, a position
$x$ in the undeformed crystal will map to a new position $x'$ in
deformed coordinates given by
\begin{equation}
\begin{split}
  x' & = x + \int_0^x \varepsilon(v) dv \\
  & = x + \frac{\sqrt{\pi} L}{8}\varepsilon_{\rm max}
  \left[\mathrm{erf} \left(\frac{x-L/2}{L/4}\right) - \mathrm{erf}(-2)
  \right],
\end{split}
\end{equation}
where $\mathrm{erf}(x)$ denotes error function.

Denote by $V(x')$ the crystal potential of the 1D crystal after
applying the Gaussian inhomogeneous strain, we then adopt the local
approximation of crystal potential as described in Sec.~\ref
{subsec:locality}, which says that the inhomogeneously strained
crystal potential at point $x'$ can be well approximated by the
crystal potential of a homogeneously strained crystal with the same
strain value. This can be mathematically written out as
\begin{equation}
V(x') \approx U^*(x; \varepsilon(x)), \quad x' = x + \int_0^x \varepsilon(v) dv.
\end{equation}
The as-constructed strained crystal potential $V(x')$ is visualized in
Fig.~\ref{fig:fig3}.

We have thus, for demonstration purpose, explicitly constructed the
strained crystal potential $V(x')$ using the local approximation of
crystal potential. This allows us to solve the energy eigenstates of
an inhomogeneously strained crystal using two distinct methods:

Method 1: direct numerical diagonalization of strained
Hamiltonian. Since the explicit expression for the inhomogeneously strained crystal
potential $V(x')$ has been constructed, we can solve the
Schr\"{o}dinger equation for the inhomogeneously strained crystal in
deformed coordinates,
\begin{equation}
\left[-\frac{1}{2}\frac{d^2}{dx'^2} + V(x') \right] \Psi(x') = E \Psi(x'),
\label{eq:1D_Deformed}
\end{equation}
by diagonalizing the Hamiltonian $H = -\frac{1}{2}\frac{d^2}{dx'^2} +
V(x')$ using plane wave basis set in Fourier space. More
straightforwardly, we can discretize the wavefunction $\Psi(x')$ into
a $N \times 1$ matrix vector in real space,
\begin{equation}
\Psi(x') = \begin{bmatrix}
\Psi(x'_1) & \Psi(x'_2) & \cdots & \Psi(x'_N)
\end{bmatrix}^T,
\end{equation}
and then write the Hamiltonian as a matrix operator $\mathbf{H}$
acting on the wavefunction $\mathbf{H} = -\mathbf{L}/2 + \mathbf{V}$,
where $\mathbf{L}$ and $\mathbf{V}$ are the matrix operators for the
differential operator $\frac{d^2}{dx'^2}$ and the potential operator
$V(x')$ respectively:
\begin{equation}
  \mathbf{L} = \frac{1}{(\Delta x')^2}
\begin{bmatrix}
  -2   &   1    &        &  1 \\
  1    &  -2    & 1      &    \\
       &   1    & \ddots &  1  \\
  1    &        & 1      & -2
\end{bmatrix},
\end{equation}
\begin{equation}
\mathbf{V} =
\begin{bmatrix}
  V(x'_1)&         &        &     \\
        &  V(x'_2) &        &     \\
        &        & \ddots &   \\
        &          &      & V(x'_N)
\end{bmatrix}.
\end{equation}
$\Delta x' = x'_{i+1} - x'_{i}$ is the distance between two real space
grid points. The Hamiltonian matrix $\mathbf{H}$ can then be
numerically diagonalized to obtain the energy eigenvalues $E$ and
wavefunctions $\Psi(x')$.

Method 2: solving the energy eigenstates of inhomogeneously strained
crystal using our envelope function method. We can solve the
Schr\"{o}dinger equation, Eq.~\ref{eq:1D_Deformed}, by first mapping
it back to undeformed coordinates, which becomes
\begin{equation}
\left[\mathcal{P}^* + U^*(x;\varepsilon(x))\right] \Psi^*(x) = E\Psi^*(x).
\label{eq:1D_undeformed_coordinates}
\end{equation}
The explicit expression for the differential operator
$\mathcal{P}^*$ is given by Eq.~\ref{eq:OperatorP_1D}.  The mapped
wavefunctions $\Psi^*(x)$ will then be expressed in terms of envelope
functions $F_n(x)$ and strain-parametrized Bloch functions $\ux$:
\begin{equation}
  \Psi^*(x) = \sum_n F_n(x) \ux.
  \label{eq:1D_envelope_expansion}
\end{equation}
We then follow the procedures described in
Sec.~\ref{sec:1DFramework} to eliminate the crystal potential term
$U^*(x;\varepsilon(x))$ in Eq.~\ref{eq:1D_undeformed_coordinates}
using strain-parametrized Bloch functions $\ux$ and the associated
strain-parametrized energy eigenvalues $\epsilon(x;
\varepsilon(x))$. Eq.~\ref{eq:1D_undeformed_coordinates} can then be
turned into a coupled differential eigenvalue equation for the
envelope functions $F_n(x)$ given by
Eq.~\ref{eq:Envelope_Function_1D}, and solved as a generalized matrix
eigenvector problem.

The solution of Eq.~\ref{eq:Envelope_Function_1D} requires the
explicit construction of strain-parametrized functions $\ux$ and the
associated strain-parametrized energy eigenvalues $\epsilon(x;
\varepsilon(x))$. The construction of these functions involves
unit-cell level calculations of homogeneously strained 1D
crystals. Only the Bloch functions and energy eigenvalues of the
electronic states at the reference crystal momentum ($k = 0$ in this
case) and a few bands close to the valence/conduction band need to be
calculated. The homogeneous strain values $\varepsilon$ are coarsely
taken from the inhomogeneous strain field (no more than one grid point
per unit cell). The calculated periodic Bloch functions of each
homogeneously strained crystal are then expressed in plane wave basis
as $u_{n0}^*(x;\varepsilon) = \sum_m C_m^n(\varepsilon) e^{i2\pi
  mx/a_0}$, where $C_m^n(\varepsilon)$ are the expansion
coefficients. The strain-parametrized Bloch functions can then be
constructed by letting $\varepsilon = \varepsilon(x)$ at position $x$,
namely
\begin{equation}
  u_{n0}^*(x; \varepsilon(x)) = \sum_m C_m^n(\varepsilon(x)) e^{i2\pi
    mx/a_0}.
\end{equation}
It is easy to see that, at position $x$, the value of $u_{n0}^*(x;
\varepsilon(x))$ is the same as the value of periodic Bloch function
$u_{n0}^*(x;\varepsilon)$ at $x$ and $\varepsilon = \varepsilon(x)$.
It is in this sense $u_{n0}^*(x; \varepsilon(x))$ are named
strain-parametrized functions.  Since $C_m^n(\varepsilon)$ are smooth
functions of $\varepsilon$ and $\varepsilon(x)$ is a smooth function
of $x$, we can use polynomial fitting to obtain smooth functions
for $C_m^n(\varepsilon(x))$. As only unit-cell level calculations of
homogeneously strained crystals at a reference crystal momentum are
involved, the construction of the strain-parametrized functions $\ux$
and $\epsilon(x; \varepsilon(x))$ do not require much computational
power in this 1D example.

\begin{figure}[t]
 \includegraphics[width=0.45\textwidth]{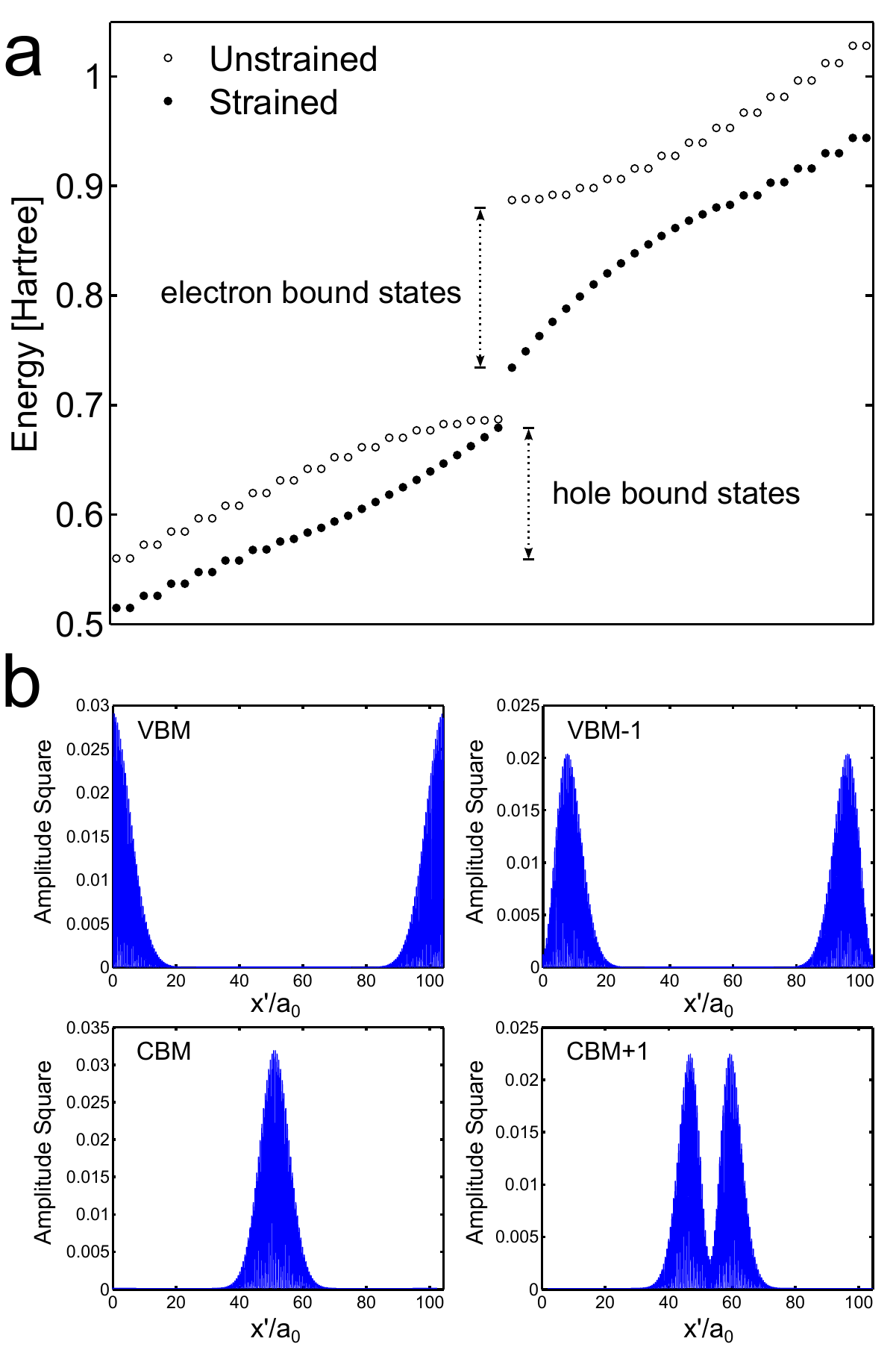}
 \caption{\label{fig:fig4} (a) Energy eigenvalues of the unstrained
   (open circles) and inhomogeneously strained model 1D crystal
   (filled circles) obtained by direct diagonalization. The energy
   levels are shifted horizontally with respect to each other to
   resolve energy levels which are very close to each other.  The
   energy range of hole and electron bound states in strained crystals
   are labeled. (b) Wavefunction probability amplitude for hole and
   electron bound states, which are labeled in the figure as VBM,
   VBM-1, CBM, and CBM+1. VBM denotes valence band maximum; VBM-1
   denotes one energy level below VBM; CBM means conduction band
   minimum, while CBM+1 denotes one energy level above CBM. The
   wavefunctions have rapid oscillation.}
\end{figure}

Of the two methods discussed above, Method 1, the direct
diagonalization of Hamiltonian, is a well established method,
therefore it can be used to benchmark Method 2, our envelope function
method. To test the effectiveness of our envelope function method, we
have calculated the energy eigenvalues and eigenfunctions of the 1D
inhomogeneously strained crystal using both methods. A special note is
that we are not testing here how good the local approximation of
crystal potential for inhomogeneously strained crystal can be, but how
accurate and fast our envelope function method can achieve given the
local approximation of crystal potential is a sufficiently good
approximation. Also note that, although for the sake of benchmarking
our envelope function method, we have explicitly constructed the
strained crystal potential in this 1D problem, in practical
application of our envelope function method, such explicit
construction of crystal potential will not be performed. The
information of local strained crystal potential, at the level of
approximation used in our method, is reflected in the
strain-parametrized Bloch functions $\ux$ and the associated
strain-parametrized energy eigenvalues
$\epsilon_{n0}(x;\varepsilon(x))$.

Choosing the following model parameters $L = 100a_0$, $U_0 =0.2$, $K =
-0.5$, and $\varepsilon_{\rm max} = 0.1$ for the 1D inhomogeneously
strained crystal, we carry out numerical real space diagonalization of
the Hamiltonian by spatially discretizing the wavefunction $\Psi(x)$
into a $N\times 1$ matrix. Periodic boundary condition $\Psi(0) =
\Psi(L)$ is adopted. As the wavefunction oscillates rapidly even
within a unit cell, very large $N$, around 50 times the number of unit
cell $L/a_0$, is needed to achieve convergence of energy eigenvalues
near valence or conduction band edge.

\begin{figure}[t]
 \includegraphics[width=0.45\textwidth]{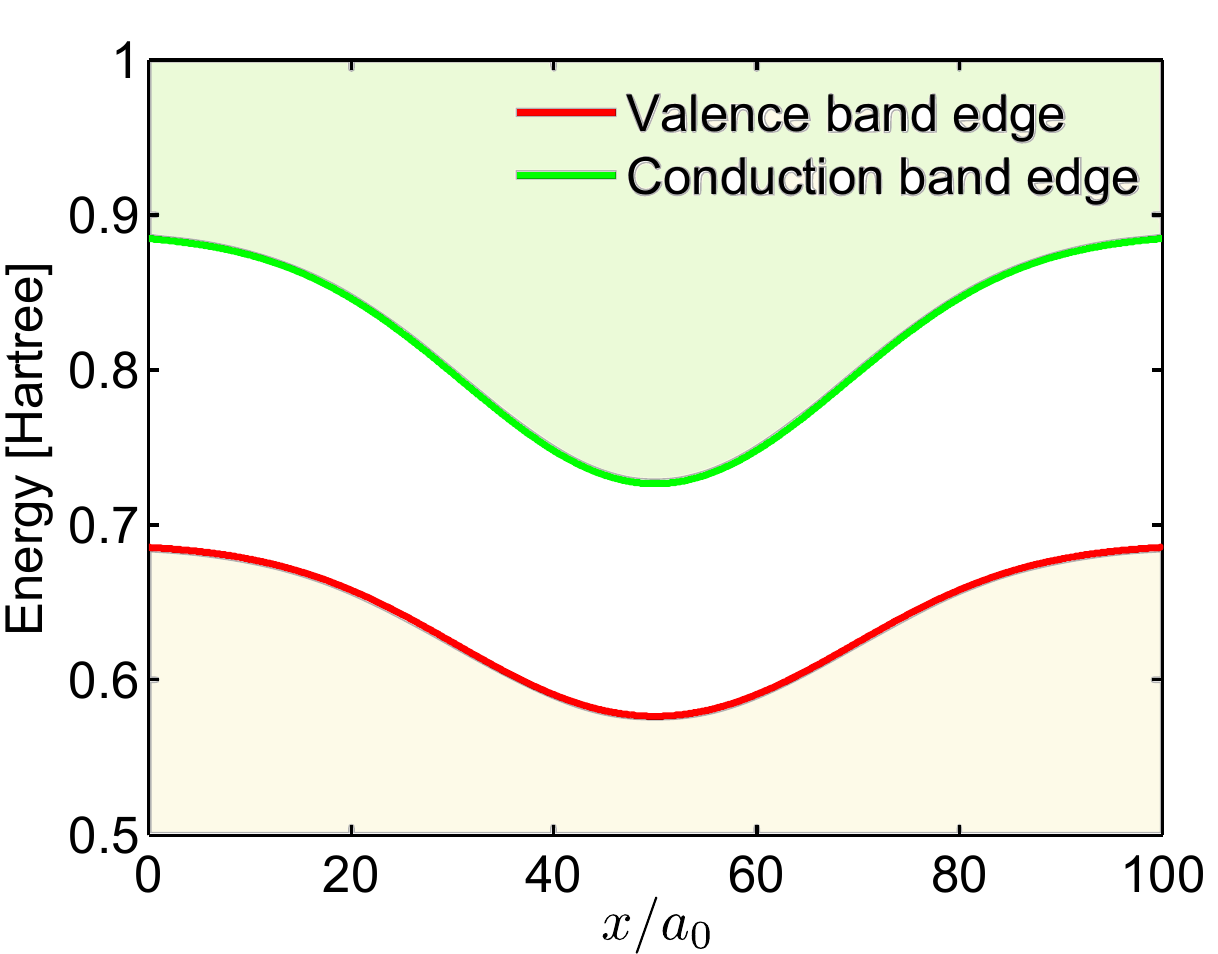}
 \caption{\label{fig:fig5} Valence and conduction band edge plotted as
   a function of position in inhomogeneously strained crystal. The
   local band edges are calculated from homogeneously strained crystal
   with the same strain magnitude at position $x$.}
\end{figure}

Fig.~\ref{fig:fig4}a shows the direct-diagonalization obtained energy
eigenvalues near the band edges.  A $5000\times 5000$ Hamiltonian matrix
is involved in the numerical calculation.  For comparison, the energy
eigenvalues of unstrained crystal are shown together in the
figure. The most distinct feature for the energy spectrum of
inhomogeneously strained crystal is the appearance of bound states
near the conduction and valence band edges. These bound states, whose
wavefunctions are shown in Fig.~\ref{fig:fig4}b, can be understood by
plotting the local valence and conduction band edges as a function of
position in the strained crystal, which is shown in
Fig.~\ref{fig:fig5}. The alignment of band edges is reminiscent of
semiconductor quantum well, except that in our case, the spatial
variation of band-edge is smooth and extended, while in semiconductor
quantum well, band edge usually jumps abruptly at the interface
between the barrier and well region of quantum well. Hence, the
strain-confined bound states in inhomogeneously strained crystal bear
resemblance to bound states in quantum well. We want to emphasize
that, the band edge alignment in our 1D inhomogeneously strained
crystal is not unique to this model. Strain-induced band edge shift in
semiconductor is a well-known phenomenon.\cite{Bardeen50} In fact, the
band-edge alignment in our 1D model is similar to those calculated by
Feng \textit{et al} for inhomogeneously strained MoS$_2$ monolayer.
\cite{Feng12} We can therefore conclude that the existence of electron
or hole bound states is a general feature in an inhomogeneously
strained crystal.

\begin{figure}[t!]
 \includegraphics[width=0.4\textwidth]{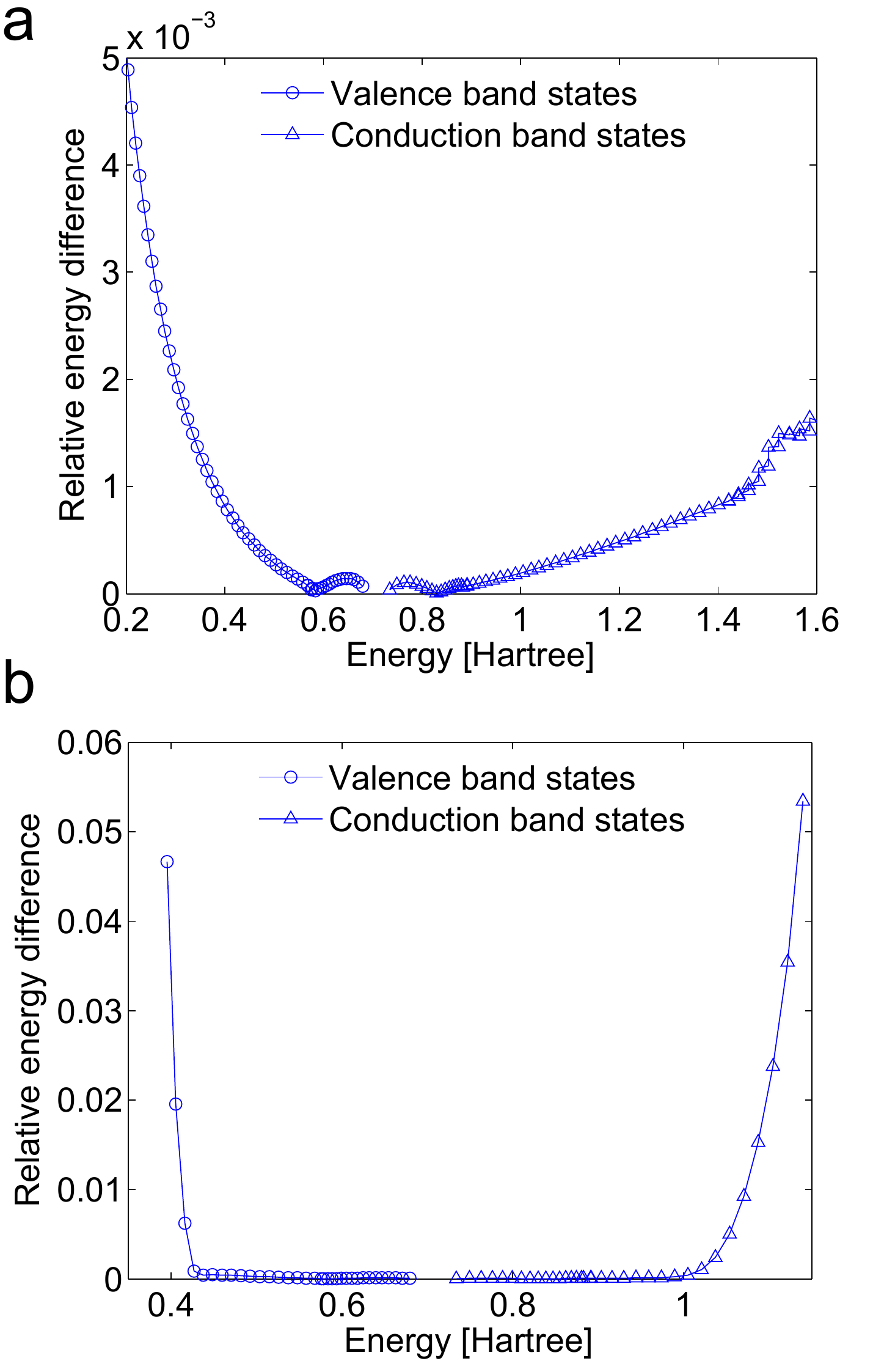}
 \caption{\label{fig:fig6} Relative difference of energy eigenvalues
   obtained by direct diagonalization and envelope function
   method. The energy eigenvalues from direct diagonalization of a
   5000 by 5000 Hamiltonian matrix are served as reference to
   calculate the relative difference. In (a), zone-center Bloch
   functions of the lowest five bands are used to carry out envelope
   function expansion. The envelope functions are represented
   numerically using one mesh grid every unit cell. This leads to the
   diagonalization of an approximately 500 by 500 matrix. In (b), only
   valence and conduction bands zone-center Bloch functions are
   involved in envelope function expansion. The envelope functions are represented
   using one mesh grid every four unit cells. The resulting
   matrix for diagonalization is of order 50 by 50.}
\end{figure}

We have also calculated the energy eigenvalues using our envelope
function method. As shown in Fig.~\ref{fig:fig6}a, very high accuracy
of eigenvalues is achieved for the whole valence and conduction bands
using only one mesh grid per unit cell representation of the envelope
functions. The lowest five bands are included in the summation over
bands in the envelope function expansion
(Eq.~\ref{eq:1D_envelope_expansion}). Together, the envelope function
method involves the diagonalization of an approximately $500$ by $500$
matrix, which is an order of magnitude smaller than direct
diagonalization. As zone-center Bloch functions are used to carry out
envelope function expansion, naturally the error for energy
eigenvalues near the band edge is smaller, same as in conventional
envelope function method. Furthermore, if one is only concerned with
energy levels near the band edge, which in most practical application
is true, the expense of envelope function method can be reduced by
another order of magnitude by including only the most relevant bands,
and using coarser grids for numerical representation of the envelope
functions.  In Fig.~\ref{fig:fig6}b, we show that more than 1/4 of
energy levels in valence and conduction bands can be calculated with
very high accuracy by including only the valence and conduction bands
in wavefunction expansion, and using one mesh grid every four unit
cells to represent the envelope functions. In this case, one ends up
diagonalizing a $50$ by $50$ matrix, which is two order of magnitude
smaller than direct diagonalization. Indeed, for this 1D model, our
envelop function method is much faster than the direct diagonalization
method.

\begin{figure}[t!] %
 \includegraphics[width=0.475\textwidth]{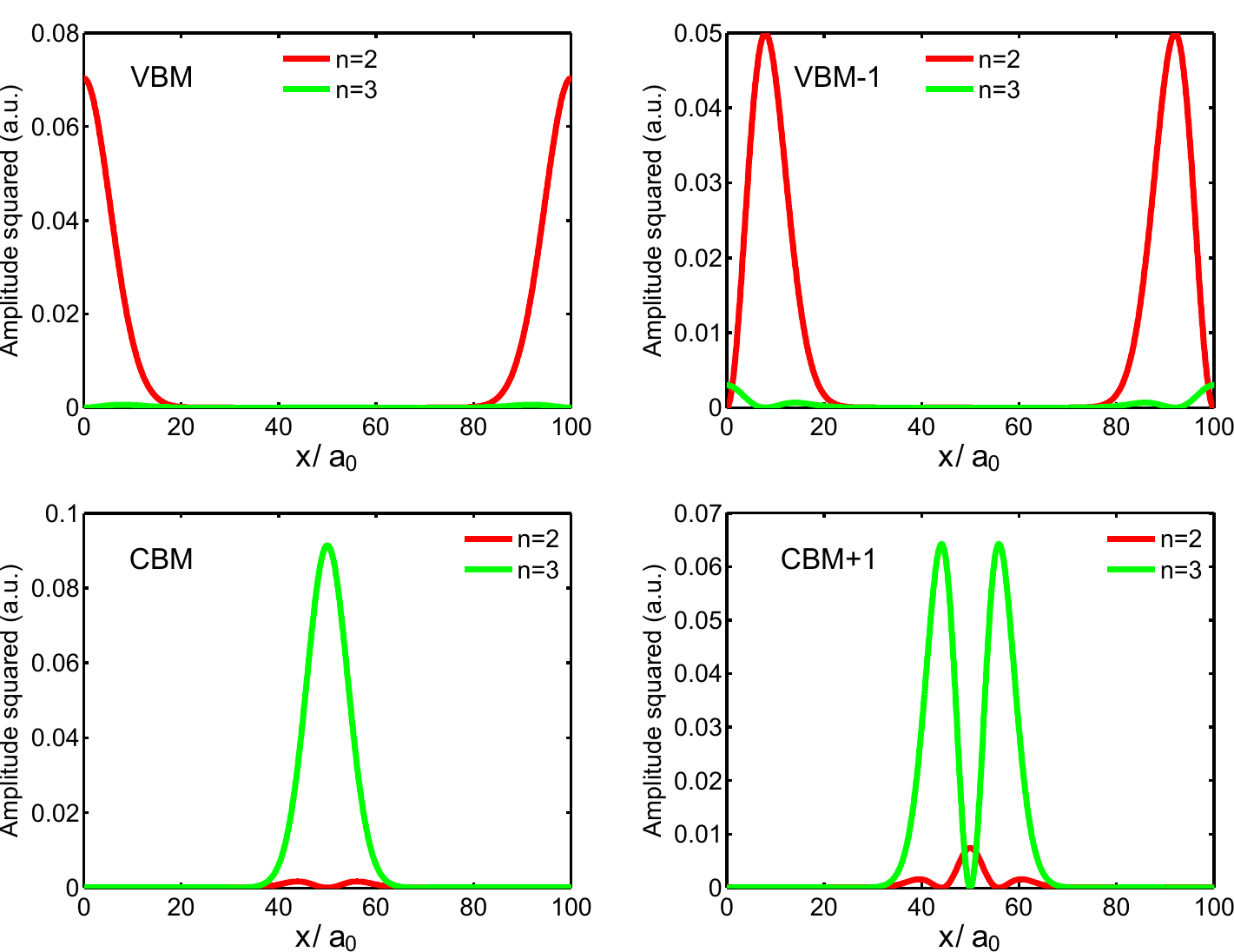} %
 \caption{\label{fig:fig7} Amplitude square plot of envelope functions
   $F_n(x)$ for states near valence and conduction band edges. The
   electronic states plotted are VBM, VBM$+1$, CBM and CBM$-1$. For
   these band edge states, only valence band ($n=2$) and conduction
   band ($n=3$) have significant envelope function amplitudes.} %
\end{figure} %

The success of the envelope function method is because the envelope
functions $F_n(x)$ are indeed slowly varying as we
conjectured. Fig.~\ref{fig:fig7} shows the amplitude square plot of
envelope functions for a few electron and hole bound states. For the
electron bound states, the envelope function of conduction band is
predominant, while the valence band envelope function also
contributes. The opposite is true for the hole bound states. Other
remote bands have negligible contribution and are therefore not
plotted.  Comparing with the full wavefunctions calculated from direct
diagonalization in Fig.~\ref{fig:fig4}b, one can notice that the
envelope functions are indeed slowly-varying functions modulating the
amplitude of fast-varying Bloch functions.

\section{Toward Application to Three-Dimensional Real
  Materials \label{sec:3DIssues}}

We have demonstrated in the previous section that our envelope
function method can be successfully applied to a model 1D
slowly-varying inhomogeneously strained semiconductor. A real
semiconductor, however, is a three-dimensional (3D) object, and its
crystal potential and strain response will be more complicated than
the 1D model. Therefore, in this section we discuss some of the issues
that may arise when applying our method to real 3D semiconductor
crystals.

\begin{figure*}[t!]
 \includegraphics[width=0.9\textwidth]{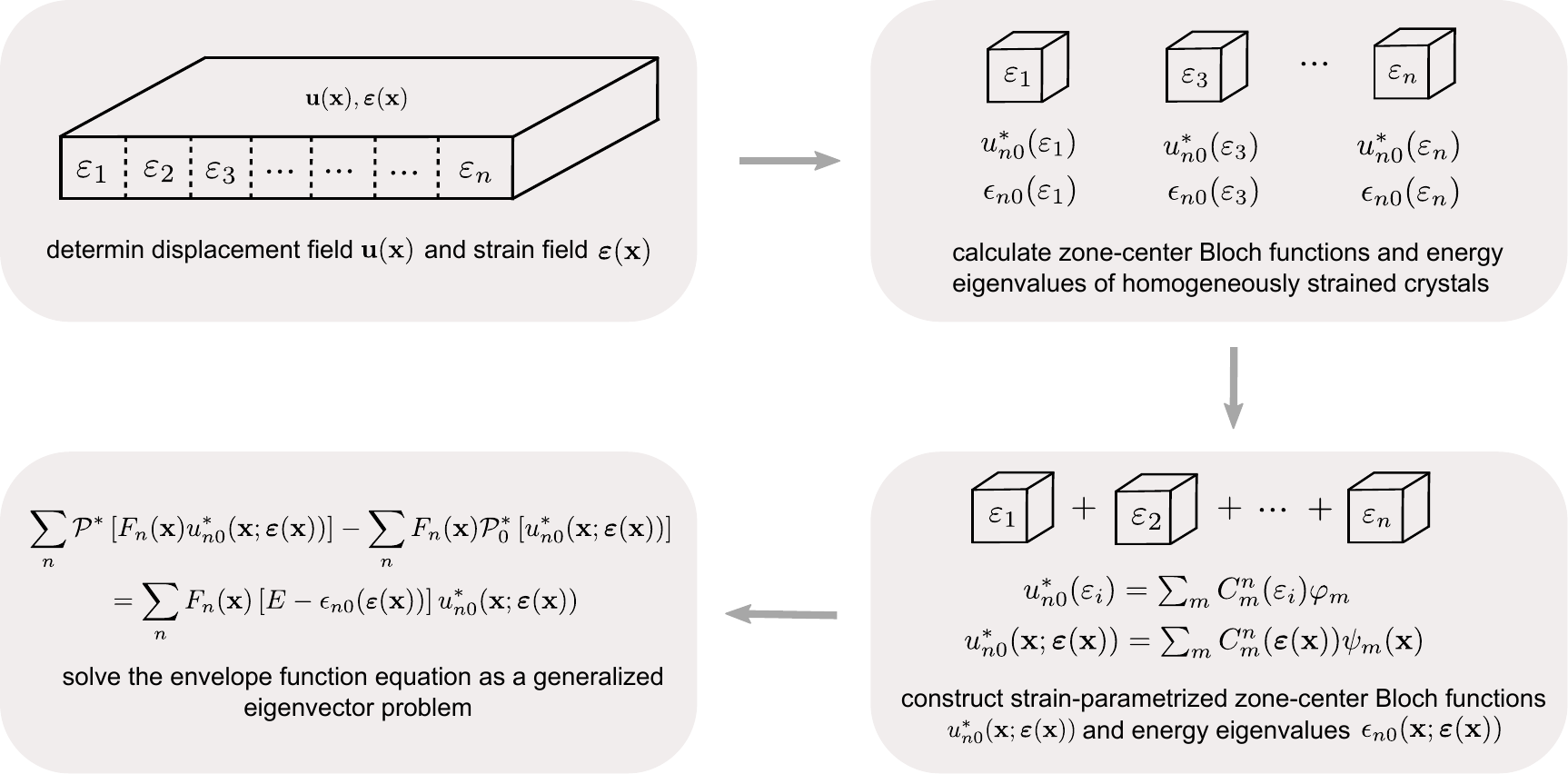}
 \caption{\label{fig:fig8} Flow chart to implement the envelope
   function method described in this article. The first step is the
   determination of a smooth displacement field $\bu(\bx)$ which can map the
   unstrained crystal (and the associated vacuum space, if any) to the
   strained crystal. The strain field $\varepsilon(\bx)$ can be
   calculated from the displacement field $\bu(\bx)$. The second and
   third steps are construction of strain-parametrized Bloch functions
   and energy eigenvalues at the reference crystal momentum (usually
   Brillouin zone center), through \textit{ab initio} or
   semi-empirical electronic structure calculations of a series of
   homogeneously strained crystal, using strain values taken from the
   inhomogeneously strained crystal. The last step is the solution of
   the coupled differential equation for the envelope functions as a
   generalized matrix eigenvector problem.}
\end{figure*}

The procedures to carry out our envelope function method in 3D are
essentially the same as in 1D, which we summarize in the flow chart
of Fig.~\ref{fig:fig8}.

The first step in the flow chart is the determination of a smooth
displacement field $\bu(\bx)$ which can map the unstrained crystal
(and the associated vacuum space, if any) to the strained crystal. The
corresponding strain field $\varepsilon(\bx)$ needs to be calculated
as well. In 1D, displacement field $u(x)$ is a one-dimensional
function and contains no rotational component. Thus $u(x)$ is related
to the strain field $\varepsilon(x)$ via a simple integral relation
$u(x) = \int^x\varepsilon(v)dv$. In 3D, the displacement field
$\bu(\bx)$ is three-dimensional, and the strain field
$\boldsymbol \varepsilon(\bx)$ is a tensor field with six independent
components. Due to the possible existence of rotational components,
the components of the strain field are related to the displacement
field $\bu(\bx)$ (in the small deformation limit) as:
\begin{equation}
  \varepsilon_{ij} = \frac{1}{2}\left(\frac{\partial u_i}{\partial x_j} + \frac{\partial u_j}{\partial x_i} \right)
\end{equation}
Hence, for a generic 3D inhomogeneously strained crystal, finding and
representing the smooth displacement field $\bu(\bx)$ and strain field
$\boldsymbol \varepsilon(\bx)$ becomes more difficult than
1D. Knowledge of solid mechanics will be helpful in this endeavor. The
increased complexity of the displacement field and strain field in 3D
also complicates the calculation and representation of the
differential operators $\mathcal{P}^*$ and $\mathcal{P}_0^*$ defined
in Eq.~\ref{eq:operator_P} and Eq.~\ref{eq:operator_P0}, which need to
be determined in order to solve the envelope function equation,
Eq.~\ref{eq:Coupled_Envelope_Function_Equation}.

The second and third step in the flow chart are the construction of
strain-parametrized Bloch functions and associated strain-parametrized
energy eigenvalues at a reference crystal momentum, usually at the
Brillouin zone center, through \textit{ab initio} or semi-empirical
electronic structure calculation of a series of \textit{homogeneously}
strained crystal. The strain values are coarsely taken from the
inhomogeneous strain field $\varepsilon$, which in principle is
sufficient as the strain field is slowly-varying in
space. Nevertheless, in a generic 3D case this step will be
challenging as the number of calculations for homogeneously strained
crystal can become quite large if the strain field is complex, as
there are six independent components of strain tensor in 3D. The
construction of strain-parameterized Bloch functions, described in
Sec.~\ref{subsec:parametrized_basis}, might also become non-trivial
due to the complexity of electronic wavefunctions in 3D. Proper choice
of expansion basis $\varphi_m(\bx)$ for the Bloch functions in
Eq.~\ref{eq:Strained_Blochwave_Expansion} and
Eq.~\ref{eq:Parametrized_Expansion_Basis} will be essential.

The fourth step in the flow chart is the solution of coupled
differential equation for the envelope functions, the
Eq.~\ref{eq:Coupled_Envelope_Function_Equation}. This step, having
been discussed in Sec.~\ref{subsec:locality}, should be straightforward
once the differential operators $\mathcal{P}^*$ and $\mathcal{P}_0^*$,
strain parameterized Bloch functions $\ubx$ and the associated strain
parameterized energy eigenvalues $\epsilon_{n0}(\varepsilon(\bx))$
have all been determined in the previous steps. Nevertheless, the
computational cost of solving the differential eigenvalue equation
will become larger as the dimensionality of the problem increases, as
more spatial or Fourier grids will be needed to represent the envelope
functions $F_n(\bx)$, resulting in larger matrices for numerical
diagonalization.

In summary, the application of our envelope function method to a
generic 3D problem will be feasible but challenging. We therefore
believe that our method will most likely find applications in cases
where the 3D problem is quasi-1D or 2D, namely when only one or very
few components of the strain tensor is varying slowly in space.

We also comment here a few issues related to the central approximation
adopted in our method, the local approximation of crystal potential in
strained crystal elaborated in Sec.~\ref{subsec:locality}. The
approximation states that in a slowly-varying inhomogeneously strained
semiconductor or insulator, the local crystal potential $V(\bx')$ can be
well approximated by that of a homogeneously strained crystal with the same
strain tensor $\varepsilon(\bx')$. This local approximation of strained
crystal potential is likely to be a good approximation only for
non-polar semiconductors such as silicon and germanium.  For polar
semiconductors such as gallium arsenide, strain could induce
piezoelectric effect, which generates long-range electric field in the
deformed crystal and significantly increases the error of this
approximation. Furthermore, we note that for certain materials with
more than one atoms within a unit cell, strain can induce internal
relaxation of atoms relative to each other on top of the displacement
described by strain tensor, an effect not included in our present
method and must be carefully checked in realistic calculations.


\section{Envelope Function Equation for Empirical
  Applications \label{sec:empirical}}
In this section, we will cast the envelope function equation
(Eq.~\ref{eq:Coupled_Envelope_Function_Equation}) in a new form in
which the strain-parametrized Bloch functions $\ubx$ will not appear
explicitly. They will be replaced by a set of matrix elements
involving their integrals. Doing so allows the method to be used
empirically, where the matrix elements can be fitted to experimental
data. The connection to traditional $\kp$ envelope function method
will also become clearer. For convenience, we rewrite the relevant
equations below
\begin{equation}
\begin{split}
 & \sum_n\mathcal{P}^*\left[F_n(\bx) \ubx \right] - \sum_n F_n(\bx) \mathcal{P}_0^*\left[\ubx\right] \\
 & =\sum_n F_n(\bx) \left[ E - \epsilon_{n0}(\varepsilon(\bx)) \right]\ubx,
\end{split}
\label{eq:Reproduced_Envelope_Function_Equation}
\end{equation}
where
\begin{eqnarray}
&&\mathcal{P}^*= -\frac{1}{2} a_{ij}(\bx)\frac{\partial ^2}{\partial x_i \partial x_j} -\frac{1}{2} b_i(\bx)\frac{\partial}{\partial x_i}, \\
&&\mathcal{P}_0^* = -\frac{1}{2}(I+\varepsilon(\bx))^{-1}_{im}(I+\varepsilon(\bx))^{-T}_{mj}
  \left.\frac{\partial ^2}{\partial x_i \partial x_j}\right|_{\varphi(\bx)}.
\end{eqnarray}
$a_{ij}(\bx)$ and $b_{i}(\bx)$ are given by
\begin{eqnarray}
  && a_{ij}(\bx) =  \left(I+\nabla \bu(\bx) \right)_{im}^{-1} \left( I+\nabla \bu(\bx) \right)_{mj}^{-T}, \\
  && b_i(\bx)   =   \left( I+\nabla \bu(\bx) \right)_{nm}^{-1} \frac{\partial}{\partial x_n} \left(I+\nabla \bu(\bx) \right)_{mi}^{-T}.
\end{eqnarray}
$\mathcal{P}^*\left[F_n(\bx) \ubx \right]$ can be expanded out as
\begin{equation}
\begin{split}
  &\mathcal{P}^*[F_n(\bx) \ubx] = \\
  & [\mathcal{P}^*F_n(\bx)] \ubx + F_n(\bx)[\mathcal{P}^* \ubx]\\
  & - a_{ij}(\bx)\frac{\partial F_n(\bx)}{\partial x_i}\frac{\partial}{\partial x_j}\ubx.
\end{split}
\end{equation}
In above expansion, we have used the symmetry property of $a_{ij}(\bx)$, namely $a_{ij}(\bx) = a_{ji}(\bx)$.

When strain variation $\varepsilon(\bx)$ is varying slowly at atomic
scale, which is the premise of our envelope function method, the
strain-parametrized basis functions $\ubx$ for different bands $n$ are
approximately independent and orthogonal:
\begin{equation}
\frac{1}{V} \int d \bx \ J(\bx) \left[\ubx \right]^{\dagger} \ubxm \approx \delta_{mn}.
\end{equation}
The integration is over the whole crystal, whose volume is $V$. The
Jacobian of deformation map $J(\bx) = \det(I + \nabla \bu)$ takes into
account the change of volume elements during coordinate
transformation. We also note that, $J(\bx)$ can be absorbed into the
basis functions by re-defining $\ubx$ as $J(\bx)^{1/2}\ubx$, and the
whole formalism of our envelope function method will not change. This
can be sometimes be more convenient for constructing
strain-parametrized basis set.

Using the above orthonormal relation, we can express $\mathcal{P}^*
\left[ \ubx \right]$, $\mathcal{P}_0^*\left[ \ubx \right]$, and
$\frac{\partial}{\partial x_i} \ubx$ in terms of $\ubx$ as
\begin{eqnarray}
&&\mathcal{P}^*   \left[ \ubx \right] = \sum_{n'} P_{nn'}\ubxp, \nonumber \\
&&\mathcal{P}_0^* \left[ \ubx \right] = \sum_{n'} P_{nn'}^0\ubxp, \nonumber \\
&&\frac{\partial}{\partial x_i} \ubx  = \sum_{n'} Q_{nn'}^i \ubxp, \nonumber
\end{eqnarray}
where $P_{nn'}$, $P_{nn'}^0$ and $Q_{nn'}^i$ are matrix elements given by
\begin{eqnarray}
&& P_{nn'}   = \frac{1}{V} \int d \bx \ J(\bx) \left\{ \mathcal{P}^*\left[ \ubx \right] \right\} \left[ \ubxp \right]^{\dagger}, \nonumber  \\
&& P_{nn'}^0 = \frac{1}{V} \int d \bx \ J(\bx) \left\{ \mathcal{P}_0^* \left[ \ubx \right] \right \} \left[ \ubxp \right]^{\dagger}, \nonumber  \\
&& Q_{nn'}^i = \frac{1}{V} \int d \bx \ J(\bx) \left\{  \frac{\partial}{\partial x_i} \ubx \right \} \left[ \ubxp \right]^{\dagger}. \nonumber
\end{eqnarray}

Eq.~\ref{eq:Reproduced_Envelope_Function_Equation} can now be written in terms of $\ubx$ as
\begin{widetext}
\begin{equation}
\sum_n \left\{ \mathcal{P}^*F_n - \sum_{n'} a_{ij}(\bx) Q_{n'n}^i \frac{\partial F_{n'}}{\partial x_j} + \sum_{n'} (P_{n'n} - P_{n'n}^0) F_{n'}   \right\} u_{n0}^* = \sum_n F_n(\bx) \left[ E - \epsilon_{n0}(\varepsilon(\bx)) \right] u_{n0}^*.
\end{equation}

Equating coefficients of $u_{n0}^*$ on both side,\cite{Burt92} we arrives at a new form of envelope function equation
\begin{equation}
-\frac{1}{2} a_{ij}(\bx) \frac{\partial^2 F_n}{\partial x_i \partial x_j} - \frac{1}{2} b_i(\bx) \frac{\partial F_n}{\partial x_i} - \sum_{n'} a_{ij}(\bx) Q_{n'n}^i \frac{\partial F_{n'}}{\partial x_j} + \sum_{n'} (P_{n'n} - P_{n'n}^0) F_{n'} + \epsilon_{n0}(\varepsilon(\bx)) F_n
= E\,F_n
\end{equation}
\end{widetext}
In the equation, $a_{ij}(\bx)$ and $b_i(\bx)$ are related to deformation mapping
and can be calculated once the displacement field $\bu(\bx)$ is
known. $Q_{n'n}^i$ and $(P_{n'n} - P_{n'n}^0)$ can be calculated
either by constructing the strain-parametrized basis set or fit
empirically to experimental data. As a sanity check, when a crystal is
undeformed, namely $\bu(\bx) = 0$, we have $a_{ij}(\bx) =
\delta_{ij}$, $b_i(\bx) = 0$, $J(\bx) = 1$, $P_{nn'} = P_{nn'}^0$,
$\epsilon_{n0}(\varepsilon(\bx)) = \epsilon_{n0}$, and the envelope
function equation will become
\begin{equation}
-\frac{1}{2} \nabla^2F_n - \sum_{n'} q_{n'n}^i \frac{\partial F_{n'}}{\partial x_i} + \epsilon_{n0} F_{n} = E\, F_n
\label{eq:EnvelopeFunction_Bulk_Crystal}
\end{equation}
with $q_{n'n}^{i}$ being
\begin{equation}
q_{n'n}^i = \frac{1}{V} \int d \bx \left[ u_{n0}(\bx) \right]^{\dagger} \frac{\partial}{\partial x_i} u_{n'0}(\bx)
\end{equation}
Eq.~\ref{eq:EnvelopeFunction_Bulk_Crystal} recovers the envelope function
equation for bulk crystals.\cite{Burt92}

\section{Summary and Conclusion}
To summarize, we have developed a new envelope function formalism for
electrons in slowly-varying inhomogeneously strained crystals. The
method expands the electronic wavefunctions in a smoothly deformed
crystal as the product of slowly varying envelope functions and
strain-parametrized Bloch functions. Assuming, with justifications,
that the local crystal potential in a smoothly deformed crystal can be
well approximated by the potential of a homogeneously deformed crystal
with the same strain value, the unknown crystal potential in
Schr\"{o}dinger equation can be replaced by the a small set of
strain-parametrized Bloch functions and the associated
strain-parametrized energy eigenvalues at a chosen crystal momentum.
Both the strain-parametrized Bloch functions and strain-parametrized
energy eigenvalues can be constructed from \textit{ab initio} or
semi-empirical electronic structure calculation of homogeneously
strained crystals at unit-cell level. The Schr\"{o}dinger equation can
then be turned into eigenvalue differential equations for the envelope
functions. Due to the slowly-varying nature of the envelope functions,
coarse spatial or fourier grids can be used to represent the envelope
functions, therefore enabling the method to deal with relatively large
systems. Compared to the traditional multi-band $\kp$ envelope
function method, our envelope function method has the advantage of
keeping unit-cell level microstructure information since the local
electronic structure information is obtained from \textit{ab initio}
or EPM calculations. Compared to the conventional EPM method, our
method uses envelope function formalism to solve the global electronic
structure, therefore has the potential to reduce the computational
cost.  The method can also be used empirically by fitting the
parameters in our derived envelope function equations to experimental
data. Our method thus provides a new route to calculate the electronic
structure of slowly-varying inhomogeneously strained crystals.

\section{Acknowledgements}
We acknowledge financial support by NSF DMR-1120901.  Computational
time on the Extreme Science and Engineering Discovery Environment
(XSEDE) under the grant number TG-DMR130038 is gratefully
acknowledged.

\bibliography{reference}
\end{document}